\documentclass[aps,pre,showpacs,preprintnumbers,twocolumn,nofootinbib,superscriptaddress]{revtex4-1}
\usepackage{geometry}                
\usepackage{graphicx}
\usepackage{amssymb}
\usepackage{dsfont}
\usepackage{epstopdf}
\usepackage{color}
\usepackage{mathtools}
\usepackage{hyperref}

\usepackage{calrsfs}

\definecolor{red}{rgb}{1,0,0}
\definecolor{blue}{rgb}{0,0,1}
\definecolor{black}{rgb}{0,0,0}

\newcommand{\p}{\partial}
\newcommand{\eq}[1]{\begin{align}#1\end{align}}

\newcommand{\ffrac}[2]{\mbox{$\frac{#1}{#2}$}}
\newcommand{\half}{\mbox{$\frac{1}{2}$}}
\newcommand{\OO}{\mathcal{O}}
\newcommand{\tr}{\mbox{tr}}

\usepackage{scalerel,stackengine}
\stackMath
\newcommand\widecheck[1]{%
\savestack{\tmpbox}{\stretchto{%
  \scaleto{%
    \scalerel*[\widthof{\ensuremath{#1}}]{\kern-.6pt\bigwedge\kern-.6pt}%
    {\rule[-\textheight/2]{1ex}{\textheight}}
  }{\textheight}%
}{0.5ex}}%
\stackon[1pt]{#1}{\scalebox{-1}{\tmpbox}}%
}

\newcommand{\tauh}{\hat{\tau}}
\newcommand{\rh}{\hat{r}}
\newcommand{\qh}{\hat{q}}

\newcommand{\Qb}{\hat{Q}}

\newcommand{\Pb}{\hat{P}}

\newcommand{\Ic}{\mathcal{I}}

\newcommand{\rv}{\vec{r}}

\newcommand{\wv}{\vec{w}}
\newcommand{\qv}{\vec{q}}

\newcommand{\zetav}{\vec{\zeta}}

\newcommand{\epsv}{\vec{\epsilon}}

\newcommand{\rvp}{\vec{r}\;'}

\newcommand{\delb}{\hat\delta}
\newcommand{\sigmab}{\hat\sigma}
\newcommand{\epsb}{\hat\epsilon}

\newcommand{\sigmabbar}{{\hat{\overline \sigma}}}
\newcommand{\sigmabar}{\overline \sigma}
\newcommand{\pbar}{\overline p}

\newcommand{\D}{{\mathcal{D}}}

\newcommand{\fsl}[1]{\ensuremath{\mathrlap{\!\not{\phantom{#1}}}#1}}
\newcommand{\sigmasbar}{\fsl{\sigmabar}}
\newcommand{\sigmas}{\fsl{\sigma}}
\newcommand{\epsbbar}{\hat{\overline{\epsilon}}}
\newcommand{\eff}{\text{eff}}

\begin{document}
\title{Renormalization of elastic quadrupoles in amorphous solids}
\author{Eric De\;Giuli}
\affiliation{Institut de Physique Th\'eorique Philippe Meyer, \'Ecole Normale Sup\'erieure, \\ PSL University, Sorbonne Universit\'es, CNRS, 75005 Paris, France}
\affiliation{Department of Physics, Ryerson University, M5B 2K3, Toronto, Canada}

\begin{abstract}
Plasticity in amorphous solids is mediated by localized quadrupolar instabilities, but the mechanism by which an amorphous solid eventually fails or melts is debated. In this work { we argue that these phenomena can be investigated in the model problem of an elastic continuum with quadrupolar defects, at finite temperature. This problem is posed and the collective behavior of the defects is analytically investigated. Using both renormalization group and field-theoretic techniques, it is found that the model has a yielding/melting transition of spinodal type. } 
\end{abstract}

\maketitle

The accepted paradigm for relaxation and flow of an amorphous solid is that of a thermally vibrating elastic medium punctuated by instability \cite{Dyre06,Maloney06a,Nicolas18,Cao18}. The former can be considered as a continuum, with discreteness relegated to heterogeneity of density, elastic moduli, or pre-stress, and an ultraviolet cutoff that corresponds to the underlying particle scale. It is often lamented that for amorphous materials, there is no simple equivalent to dislocations and disclinations that govern the melting of a crystal. However, it is now well established that flow of amorphous solids is mediated by localized quadrupolar instabilities of Eshelby type \cite{Eshelby57}, and it has been argued that relaxation of a supercooled liquid can also be understood in this framework \cite{Dyre99,Lemaitre14,Lemaitre15,Buchenau18,Buchenau18a,Buchenau19,Buchenau19a}. The role of alternative mechanisms of relaxation, and the precise relation between structural heterogeneity and the location of incipient instabilities is still debated \cite{Berthier11b,Manning11,Patinet16,Gartner16,Zylberg17}, but meanwhile the importance of localized instabilities as excitations of an otherwise elastic medium is clear. Numerically, localized forcing has been used as a probe of glass properties \cite{Lerner18,Rainone19}. However, the collective behavior of many localized quadrupolar instabilities has hardly been analytically investigated. Crucial first steps were performed in \cite{Dasgupta12,Dasgupta13a}, where it was shown that in the presence of external shear stress, it is energetically favorable to align quadrupoles collinearly. This was interpreted as a precursor to the formation of macroscopic shear bands. 

In \cite{Dasgupta12}, and in some subsequent works \cite{Moshe15,Moshe15a}, the elastic self-energy of the quadrupoles was neglected, while in \cite{Dasgupta13a} it appears in a calculation of the yield strain for a line of quadrupoles. In this paper we show that this self-energy plays a crucial role in the collective behavior, even in the absence of external stress. Using methods developed by Kosterlitz, Thouless, Halperin, Nelson, and Young for the theory of 2D melting \cite{Kosterlitz73,Kosterlitz74,Nelson79,Young79}, we compute the renormalization of elastic interactions by a small density of quadrupolar defects in a two-dimensional elastic continuum. We will show that interactions can reduce the self-energy to such an extent that a shear stiffness can vanish, thus signalling a phase transition. Under external stress, we interpret this transition as the yielding of an amorphous solid, while in the absence of stress it corresponds to melting. The transitions are predicted to be continuously related, although yielding is much more abrupt than melting. 

This paper is organized as follows. First, we discuss elementary excitations of an amorphous solid in general and argue that these excitations will have a non-trivial renormalization. Then, we pose the equilibrium problem of a collection of quadrupolar defects in an elastic continuum. The corresponding partition function is then analyzed, first by a renormalization group method, and then by field-theoretic methods. Both techniques lead to the conclusion that such a solid will have a melting/yielding transition of spinodal type. We then outline how our results can be applied to out-of-equilibrium and athermal amorphous solids, and discuss prospects for future work.

Our tensor notation is such that all contractions are explicitly indicated. We alternatively use index-free notation, when appropriate, and indices when necessary, with the Einstein convention. The identity tensor is denoted $\delb$. We make use of the antisymmetric tensor, $\epsilon_{12}=-\epsilon_{21}=1, \epsilon_{11}=\epsilon_{22}=0$.


{\bf Elementary excitations: } We consider amorphous solids that can be treated as low-temperature continua. Since a solid must break translational symmetry, we are tacitly assuming that the stress field has long-range correlations \cite{Bi15,Sarkar13}, which are indeed easily accounted for in the framework \cite{DeGiuli18a}. We work in a dual description developed by Kleinert that uses stress as the fundamental variable \cite{Kleinert89a}. In two dimensions, the stress tensor can be written in terms of a scalar gauge field, the Airy stress function $\psi$, as $\sigma_{ik} = \epsilon_{ij} \epsilon_{kl} \p_j \p_l \psi$. Any configuration of $\psi(\rv)$ identically describes stress fields in mechanical equilibrium, called inherent states. The curvature of $\psi(\rv)$ determines the stress\footnote{See \cite{DeGiuli18} for a discussion of gauge freedoms.}. Since the continuum is an idealization of a collection of discrete particles, the $\psi$ field can be punctured at any point, creating defects. What type of defects are permitted? While one might imagine that $\psi$ could be multi-valued, in fact an explicit construction at the particle scale shows that the $\psi$ field is continuous at the smallest scale at which it can be defined \cite{DeGiuli14a}; at most it can have point singularities, living in the voids at the particle scale.
Their form can be motivated physically. 
 Indeed, since any elementary excitation taking one inherent state to another must preserve force and torque balance, the most basic excitation is the stress response to a dipole of forces, that is  
a pair of equal and opposite forces $\pm\vec{f}$ at a separation $\vec{s} \propto \vec{f}$, which respects both constraints. In the far-field limit $s/r \ll 1$ the change in $\psi$ due to imposed external forces $\pm \vec{f}$ at $\mp \vec{s}$ is \cite{Sokolnikoff56}
\eq{ \label{dipole1}
D(\rv; \tau, \theta) & = a_0 \tau \log r - a_2 \tau \cos(2\phi - 2\theta) \notag \\
& \qquad + \OO(\tau s^2/r^2)
}
where $a_0 = (3-\nu)/(4\pi), a_2 = (1+\nu)/(4\pi)$, $\rv = r (\cos \phi,\sin \phi), \vec{f} = f (\cos \theta,\sin \theta)$, $\tau = \vec{f} \cdot \vec{s}$ is the dipole moment of the excitation, and $\nu$ is Poisson's ratio\footnote{The Poisson ratio is related to the Lam\'e modulus by $\lambda=2\mu\nu/(1-(d-1)\nu)$ in $d$ dimensions \cite{Kleinert89a}.}. 
In the taxonomy of \cite{Moshe15a}, the first term in \eqref{dipole1} is monopolar, and the second term is quadrupolar. As pointed out in \cite{Moshe15a}, such a force dipole is actually not a local excitation. Indeed, if the locus of the defect is removed by creating a void, then the material cannot relax to a strain-free configuration; this is due to the monopolar term in \eqref{dipole1}. It is easily seen that if we add a second force dipole at an angle of $\pi/2$ with respect to the original, and with opposite sign, the monopolar terms cancel, while the quadrupolar terms add. This force quadrupole is a local excitation, and can thus be produced physically by localized instabilities. The Eshelby inclusion procedure \cite{Eshelby57} can be considered as an explicit physical realization of quadrupolar instability, but the far-field behavior is universal. In our treatment we will consider the quadrupoles as having a core radius $a$; its initial value is arbitrary so long as $a \gtrsim s$, and eventually will be renormalized away. 

{ Note that in treating the solid as an elastic continuum, we assume the validity of linear elasticity up to a wavenumber cutoff $\Lambda$, associated to the inverse of a particle length scale. Self-consistency requires that the defect core size is larger than this length scale. Fits of quadrupolar instabilities to the Eshelby inclusion procedure for a Lennard-Jones glass inferred a core involving approximately 20 particles \cite{Dasgupta13a}, indeed much larger than the size of a single particle. However, as the jamming point is approached, continuum elasticity breaks down \cite{Lerner14}; we thus need to assume that our solid is deep in the jammed phase. This is discussed further in the conclusion. }




\begin{figure}[t!]
\includegraphics[width=\columnwidth]{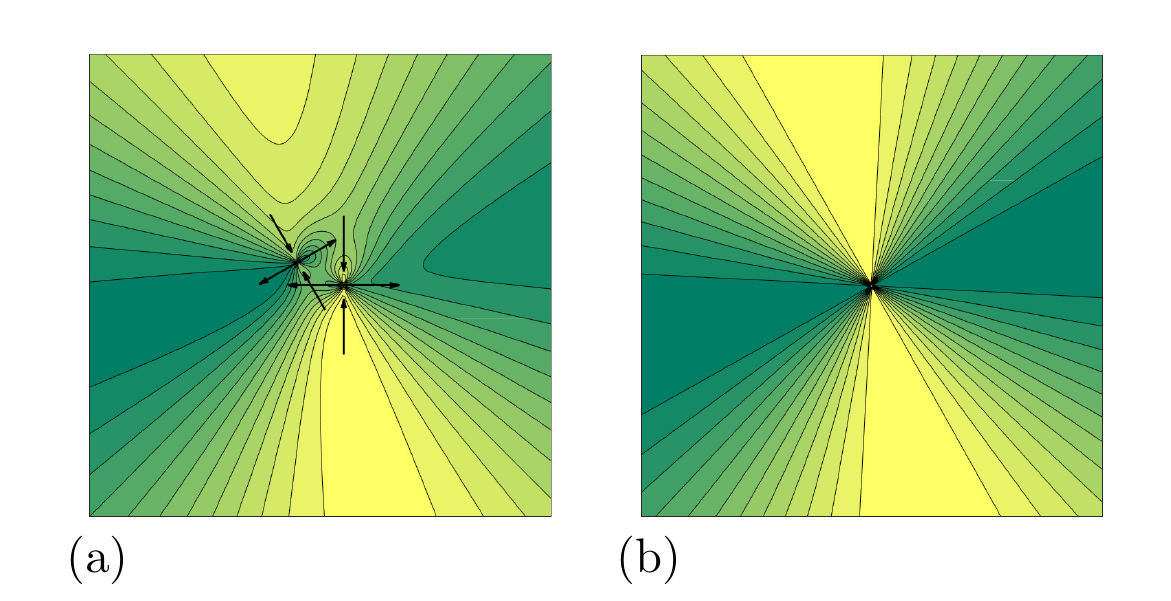}
\caption{ Airy stress function for two elastic quadrupoles (a) and a single quadrupole with the same dipole moment, (b). The fields are comparable beyond a distance $r \approx 2s$, where $s$ is their separation.
}\label{fig1}
\end{figure}

Consider two quadrupoles $\tau_1$ and $\tau_2$ at a distance $\rv$, as shown in Fig. \ref{fig1}a. They have an elastic interaction energy \cite{Moshe15a}
\eq{
U = \frac{2 a_2}{\mu} \frac{\tau_1\tau_2}{r^2} \cos(2\theta_1 + 2\theta_2 - 4 \phi)
}
where $\mu$ is the shear modulus. They also have self-energies of the form
\eq{
E_{i} = \frac{c}{\mu} \frac{\tau_i^2}{a^2} 
} 
where the coupling constant $c$ depends on the regularization at the core scale. The interaction energy is minimized when $\phi = (\pi+\theta_1+\theta_2)/2$ and the quadrupoles are close together, $r = 2a$. For simplicity, let $\tau_1=\tau_2=\tau$. At large distances, the minimal-energy state of the quadrupoles behaves as a renormalized quadrupole of moment $\tau'=2\tau$ and core radius $a'=2a$. We can define a renormalized self energy by $E'=\half (E_1+E_2+U)$, where the factor of $1/2$ ensures that the energy is invariant in the absence of interactions. 
This relation implies a renormalization of the coupling $c$ via
\eq{
\frac{c'}{\mu} \frac{\tau'^2}{a'^2} \equiv E' = \half \frac{1}{\mu} \frac{\tau^2}{a^2} \left[ 2c - \ffrac{a_2}{2} \right],
}
or $c' = c-a_2/4$, assuming that $\nu$ and $\mu$ remain invariant. 
 The elastic interaction reduces the coupling, opening the possibility that under repeated renormalization there is a non-trivial fixed point, implying scale invariance, or for the self-coupling to vanish, implying macroscopic instability. This is true even in the absence of external stress, which further favors the quadrupoles to co-align and thus behave as composite objects \cite{Dasgupta12,Dasgupta13a}. 

In this simple argument we are ignoring the distribution of $\tau$ and fluctuations in $\rv$, external stress, renormalization of $\mu$ and $\nu$, and deviations of the composite object from a true quadrupole. Most importantly, the microscopic self-energy is clearly dependent on details at the core scale. For these reasons, in the next section we elevate this computation to a renormalization group analysis where microscopic details can be forgotten. We will find, eventually, that generically the self-coupling vanishes at large enough scale, and implies instability of spinodal type. 

{\bf Renormalization: } 
A quadrupole is a bound state of a dilatant dipole ($\tau >0$) and a compressive dipole ($\tau <0$).
 We introduce the tensorial dipole moment $\tauh$, with units of stress$\times$volume. For a single force dipole $\tauh = \vec{f} \vec{s}$ it takes the form
\eq{ \label{tau1}
\hat{\tau} = \half \tau \begin{bmatrix} 1+\cos 2\theta & \sin 2\theta \\ \sin 2\theta & 1-\cos 2\theta \\ \end{bmatrix}
}
while for a quadrupole the isotropic component is absent\footnote{For a quadrupole, the second dipole has its force vector rotated by $\pi/2$ and its separation vector rotated by $-\pi/2$. Hence $\tauh = \vec{f} \vec{s} + \epsb \cdot \vec{f} \vec{s} \cdot \epsb = 2 \vec{f} \vec{s} - \tau \delb$ where we used that $\vec{f} \propto \vec{s}$.}. Introducing also 
%
%
a spatial coupling matrix
\eq{ \label{P1}
\quad \hat{P}(\rv) = \begin{bmatrix} \cos 2\phi & \;\;\sin 2\phi \\ \sin 2\phi & -\cos 2\phi \\ \end{bmatrix},
}
the expression $\tau \cos(2\phi - 2\theta)$ in \eqref{dipole1} can be written as $\tauh : \hat{P} = \tau_{ij} P_{ij}$, which is linear in the charge and therefore behaves well under renormalization. This indicates that tensorial charges are the correct level of description \cite{Lemaitre14}, so we promote the theory to one of general symmetric tensorial charges $\tauh$. 

\begin{figure}[t!]
\includegraphics[width=\columnwidth]{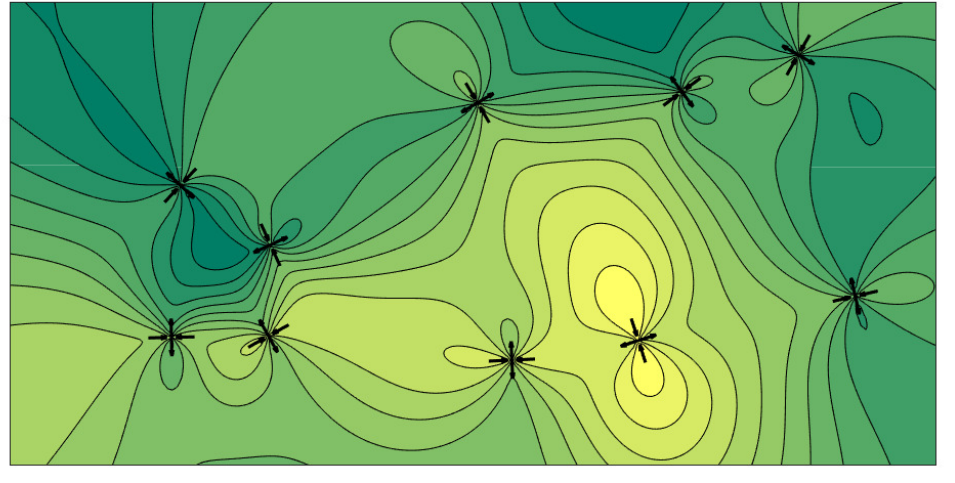}
\caption{ Airy stress function for a configuration of several quadrupolar defects. The equilibrium theory considers all such configurations, along with their phonon-mediated interactions.
}\label{manydefects}
\end{figure}


\begin{table}
  \centering
\begin{tabular}{l | p{2.5 cm} | p{4 cm} }
Symbol              & Definition & Interpretation  \\
\hline
$\tau$ & $\vec{f} \cdot \vec{s}$ & dipole moment \\
$\theta$ & & dipole orientation angle \\
$\tau_{ij}$ & Eq.\eqref{tau1} & tensorial dipole moment\\
$\vec{f}$ & & contact force \\
$\vec{s}$ & & separation vector \\
$a$ & & core radius \\
$\mu$ & & shear modulus\\
$\nu$ & & Poisson's ratio\\
$a_0$ & $(3-\nu)/(4\pi)$ & \\
$a_2$ & $(1+\nu)/(4\pi)$ & \\
$\gamma_{ij},\lambda_{ij}$ & Eq.\eqref{C1} & renormalized coupling constants \\
$\gamma^0_{ij},\lambda^0_{ij}$ & Eq.\eqref{C1} & bare coupling constants \\
$k_1,k_2$ & Eq.\eqref{V2} & self-coupling constants \\
$k^0_1,k^0_2$ & Eq.\eqref{V1} & bare self-coupling constants \\
$h$ & Eq.\eqref{V2} & pressure coupling \\
$j$ & Eq.\eqref{V2} & shear-stress coupling \\
$\Lambda$ & & wavenumber cutoff \\
$\overline{p}$ & $\frac{1}{2} \sigmabar_{kk}$ & pressure \\
$\sigmasbar_{ij}$ & $\sigmabar_{ij} - \overline{p} \delta_{ij}$ & deviatoric stress tensor \\
$\sigmabar$ & $\sqrt{\frac{1}{2}\sigmasbar_{ij}\sigmasbar_{ij}}$ & shear stress \\
$\tau_c$ & Eq.\eqref{Z1} & dipole-moment scale \\
$Z_\tau$ & Eq.\eqref{Ztau} & single defect partition function \\
$\tau_1$ & Eq.\eqref{fluc1} & average dipole-moment \\
$\tau_2$ & Eq.\eqref{fluc2} & dipole-moment fluctuation scale \\
$\epsilon_1$ & Eq.\eqref{eps1} & strain scale \\
$X$ & Eq.\eqref{X} & \\
$C_2$ & $2\pi^2a^2/(5\beta \tau_c^2)$ & \\
\hline
$\vec{\zeta},\vec{\epsilon}$ & & plastic strain fields \\
$A,B,\alpha,\tilde\alpha,\gamma$ & Eq.\eqref{couplingvector} & couplings in vectorized form \\ 
$b(q)$ & $2B \log(q/\Lambda)$ & \\
$c$ & $\beta a^2/(2k_2)$ & \\
\hline
\end{tabular}
\caption{Table of symbols. First block: symbols used in RG; second block: symbols used in field theory. } \label{tab}
\end{table}

The elastic energy \cite{Kleinert89a}
\eq{ \label{H1}
H = \frac{1}{4\mu} \int d^2 r  \left[ \sigma_{ij} \sigma_{ij} - \frac{\nu}{1+\nu} \sigma_{ii} \sigma_{jj} \right]
}
depends on the total stress $\sigma_{ij} = \sigmabar_{ij} + \sigma_{ij}^{D} + \sigma_{ij}^{P}$ decomposed into a constant component, defects, and transverse phonons\footnote{Beginning from the standard representation in terms of displacements, the stress tensor is introduced by a Hubbard-Stratonovich transformation, and the longitudinal phonons are integrated out when the field equation $\nabla \cdot \sigmab=0$ is imposed \cite{Kleinert89a}.}. The relationship between $\sigmabbar$ and the expected stress $\langle \sigmab\rangle $ is nontrivial, and will be discussed below. The defects take the form
\eq{
\sigma_{ij}^{D} = (\nabla \times \nabla \times)_{ij} \sum_{a=1}^n D(\rv-\rv_a; \tau_a, \theta_a)
}
where the double-curl operator is $(\nabla \times \nabla \times)_{ij} = (\epsb \cdot \nabla \nabla \cdot \epsb^t)_{ij} = \epsilon_{ik} \epsilon_{jl} \p_k \p_l$ and $D$ is as in \eqref{dipole1}. { Fig.\ref{manydefects} shows an example of the Airy stress function for several quadrupolar defects, that is $\sum_a D(\rv-\rv_a; \tau_a, \theta_a)$. }

As shown in Appendix 1, the transverse phonons can be integrated out to obtain the effective Hamiltonian of the defects:
\eq{ \label{H2}
H_{n'} = \sum_{a,b = 1, a\neq b}^{n'} \tau_{ij}^a \tau_{kl}^b \frac{C^0_{ijkl}(\rv_{ab})}{r_{ab}^2} + \sum_{a=1}^{n'} V^0(\tau^a_{ij}),
}
where $\rv_{ab} = \rv_a-\rv_b$. The defects have a long-range phonon-mediated interaction. We find
\eq{ \label{C1}
C^0_{ijkl} & = \gamma^0_{11} \delta_{ij} \delta_{kl} + 2 \gamma^0_{12} \delta_{ij} P_{kl} + \gamma^0_{22} P_{ij} P_{kl}  \\
& \quad + \left[ \lambda^0_{11} \delta_{ik} \delta_{jl} + 2 \lambda^0_{12} \delta_{ik} P_{jl} + \lambda^0_{22} P_{ik} P_{jl} + ( k \leftrightarrow l ) \right] \notag
}
in terms of $\Pb(\rv)$ introduced above. The $\gamma_{ij}$ and $\lambda_{ij}$ are functions of $r\Lambda$ with a constant part and a fluctuating part, where $\Lambda$ is the UV cutoff for the phonons. For simplicity, in this work we keep only the constant part, thus giving a scale-free $1/r^2$ interaction between defects, as used in most elasto-plastic models \cite{Nicolas18}. In this case we have  
\eq{ \label{init1}
\gamma^0_{22} & = 8 a_2/\mu \\
\gamma^0_{12} & = 6 (a_0-a_2)/\mu,
}
while the remaining couplings are obtained from 
\eq{ \label{init2}
\lambda_{22}^0 & = 0 \\
\lambda^0_{11} & = +\ffrac{1}{4} \gamma^0_{22} \\
\lambda^0_{12} & = +\ffrac{3}{4} \gamma_{22}^0 \\
\gamma^0_{11} & = -\ffrac{1}{4} \gamma_{22}^0. \label{init2end}
}
The local potential has the form
\eq{ \label{V1}
V^0(\tau_{ij}) & = \mu^{-1} \left[ (a_0/a_2) \pbar \delta_{ij} -\sigmasbar_{ij} \right] \tau_{ij} \notag \\
& \qquad + \ffrac{1}{2a^2} \left[ k^0_1 \delta_{ij} \delta_{kl} + k^0_2 \delta_{ik} \delta_{jl} \right] \tau_{ij} \tau_{kl}
}
where $\pbar = \half \sigmabar_{ii}$ is the pressure and $\sigmasbar_{ij} = \sigmabar_{ij} - \pbar \delta_{ij}$ is the deviatoric stress. The couplings $k_1^0$ and $k_2^0$ are not well constrained in a continuum theory, but are expected to behave as $k_i^0 \propto 1/\mu$ with an $\OO(1)$ coefficient; see Appendix 1.

Eq. \eqref{H2} applies for any set of defects of the form given in \eqref{dipole1}. We consider that we have $n'=2n$ force dipoles strictly paired into quadrupoles as above. Then the partition function for the defects is
\eq{ \label{Z1}
Z = \sum_{n \geq 0} \frac{1}{n!} \int_{r_1,\ldots,r_{n}} \int_{\tau_1,\ldots,\tau_{n}} e^{-\beta H_{2n}}
}
where $\int_{r_i}~=~\int d^2 r_i/a^2$ and $\int_{\tau_i}~=~\int d\tau^i_{xx} \int d\tau^i_{xy} \int d\tau^i_{yy} \omega[\tauh^i]/\tau_c^3$ in terms of the core radius $a$ and characteristic dipole moment $\tau_c$. The measure factor $\omega[\tauh]$ is used to enforce the correct form of the charge $\tauh$: $\omega[\tauh]=\delta(\tau_{xx}+\tau_{yy})\tau_c$ eliminates the monopolar degree of freedom.
Notice that the scale $\tau_c$ controls the fugacity of defects. We consider it as a parameter set by the quenching process from the melt. 




We aim to compute $Z$, or at least to extract the phase diagram that it describes. We will use the renormalization group in the manner of Jos\'e et al \cite{Jose77}: we consider the interaction between two fixed charges at separation $r$ and compute its renormalization by a test charge, which is integrated over. By considering an appropriate class of theories, the resulting RG equation can be transformed into an RG flow that can be iterated. 

The class of theories specified by the form of interactions must be closed under the RG. It will be sufficient to consider $H=\sum_a V(\tauh_a) + \sum_{a\neq b} U(\tauh_a,\tauh_b,\rv_{ab})$ with
\eq{ \label{U2}
U(\tauh_a,\tauh_b,\rv_{ab}) = \tau^a_{ij} \tau^b_{kl} C_{ijkl}(\rv_{ab})/r_{ab}^2
}
where $C_{ijkl}(\rv)$ is of the form \eqref{C1}, without the superscripts. The charges have quadratic self-interactions
\eq{ \label{V2}
V(\tauh) & = h \pbar \tau_{ii} + j \sigmasbar_{ij} \tau_{ij} \notag \\
& \qquad + \ffrac{1}{2a^2} \left[ k_1 \delta_{ij} \delta_{kl} + k_2 \delta_{ik} \delta_{jl} \right] \tau_{ij} \tau_{kl}.
}
In the interest of future work, this class of theories allows both dipoles and quadrupoles. The parameters $h, k_1, \gamma_{11}$ and $\gamma_{12}$ are not relevant for quadrupoles and eventually will be ignored.

The RG computation is explained in Appendix 2. We find that the RG is indeed closed if (i) we take the far-field limit, $r \gg a$, and (ii) the self-energy is much larger than the interaction energy. The latter condition is equivalent to a standard small-fugacity condition. In the case of an external shear stress, we only include the most relevant anisotropic terms, namely those affecting $h$ and $j$. 

 The computation implies that dipole moments scale as $\tau \sim a$, up to anomalous corrections, in agreement with the simple argument presented previously; this generalizes to $\tau \sim a^{d/2}$ in $d$ dimensions. 

The final result for the running in $t= \log a$ is:
 \eq{ \label{RGx}
 (a^2\pbar/\tau_c) \p_t h & = 2\pi Z_\tau  (\gamma_{11} + 2\lambda_{11} + \ffrac{1}{4} \lambda_{22}) (\tau_1/\tau_c) \\
(a^2 \sigmabar/\tau_c) \p_t j & = \ffrac{\pi}{8} Z_\tau \gamma_{22} (a_\tau/\tau_c) \label{RGx2} \\
 \p_t k_{1} & = A (Y_1-\ffrac{\pi}{4} \chi Y_2) \\
 \p_t k_{2} & = A \chi (\half\pi Y_2 + 16 \pi Y_3 ) \\
 \p_t \gamma_{11} & = A Y_4 + 2\pi A \chi \gamma_{22} \gamma_{11} \\
 \p_t \gamma_{12} & = A Y_5 + 2\pi A \chi \gamma_{22} \gamma_{12} \\
 \p_t \gamma_{22} & = 2\pi A \chi \gamma_{22}^2 \\
 \p_t \lambda_{11} & = 2\pi A \chi \gamma_{22} \lambda_{11} \\
 \p_t \lambda_{12} & = 2\pi A \chi \gamma_{22} \lambda_{12} \\
 \p_t \lambda_{22} & = 2\pi A \chi \gamma_{22} \lambda_{22}
 }
where $A=-\beta Z_\tau \tau_2/(8a^2)$, and the $Y_i$ are functions of the $\lambda_{ij}, \gamma_{ij}$ and $\chi$, given in Appendix 2. 
The stresses scale as $\sigmabar \sim \tau_c/a^2$, as expected from dimensional analysis. In fact, all the terms in parentheses in Eqs.\eqref{RGx},\eqref{RGx2} scale as $a^0$, hence the flow is homogeneous, which implies that the initial value of $a$ is forgotten and the universality hypothesis is verified. A key role is played by the single-defect partition function $Z_\tau = \int_{\tau} e^{-\beta V(\tauh)}$, which controls the dipole-moment scales and fluctuations appearing as $\tau_1, a_\tau, \tau_2,$ and $\chi$ above. This depends on the measure for the defects, which can be more general than described above. 


Before specializing to the case of quadrupoles, let us note that the linear equations Eqs.\eqref{init2}-\eqref{init2end} satisfied by the initial values of the couplings are all preserved by the RG flow. The evolution therefore takes place in a proper subset of the coupling space. 

First, we look for fixed points. Assuming $A \neq 0$ and $\chi \neq 0$ as we will check later, the flow equation for $\gamma_{22}$ requires that $\gamma_{22}=0$. This then implies that all the other interactions are stationary, and only $\gamma_{12}$ can be nonzero. 
Stationarity of $k_2$ requires either $\chi=-1$, which holds for quadrupoles, or $\gamma_{12}=\gamma_{22}$. In either case the fixed point is non-interacting. The defect partition function in this case is
\eq{ \label{ZNI}
Z_{NI} = \sum_{n \geq 0} \frac{1}{n!} \left( \int_r Z_\tau \right)^{\!n} = \exp\left( \frac{\Omega}{a^2} Z_\tau \right)
}
where $\Omega$ is the area of the system, and we ignore any steric constraints on the defects. The mean number of defects is
\eq{
\langle n \rangle = -\ffrac{1}{m} \frac{\p \log Z}{\p \log \tau_c} = \frac{\Omega}{a^2} Z_\tau,
}
{ where $\langle \cdot \rangle$ denotes an average over the Boltzmann distribution,} $m$ is the exponent in $\int_{\tau} \sim \tau_c^{-m}$ (for quadrupoles $m=2$), and the second relation holds only for \eqref{ZNI}. In this approximation, $Z_\tau$ is the average number of defects in a region of area $a^2$.

Since the only fixed point is non-interacting, this suggests that the model is trivial. However, by analyzing the RG flow, we will see that this fixed point is not necessarily reached. Instead, at large enough scale there is a regime where some fluctuations diverge, signalling a proliferation of quadrupoles. 

\begin{figure}[t!]
\includegraphics[width=\columnwidth]{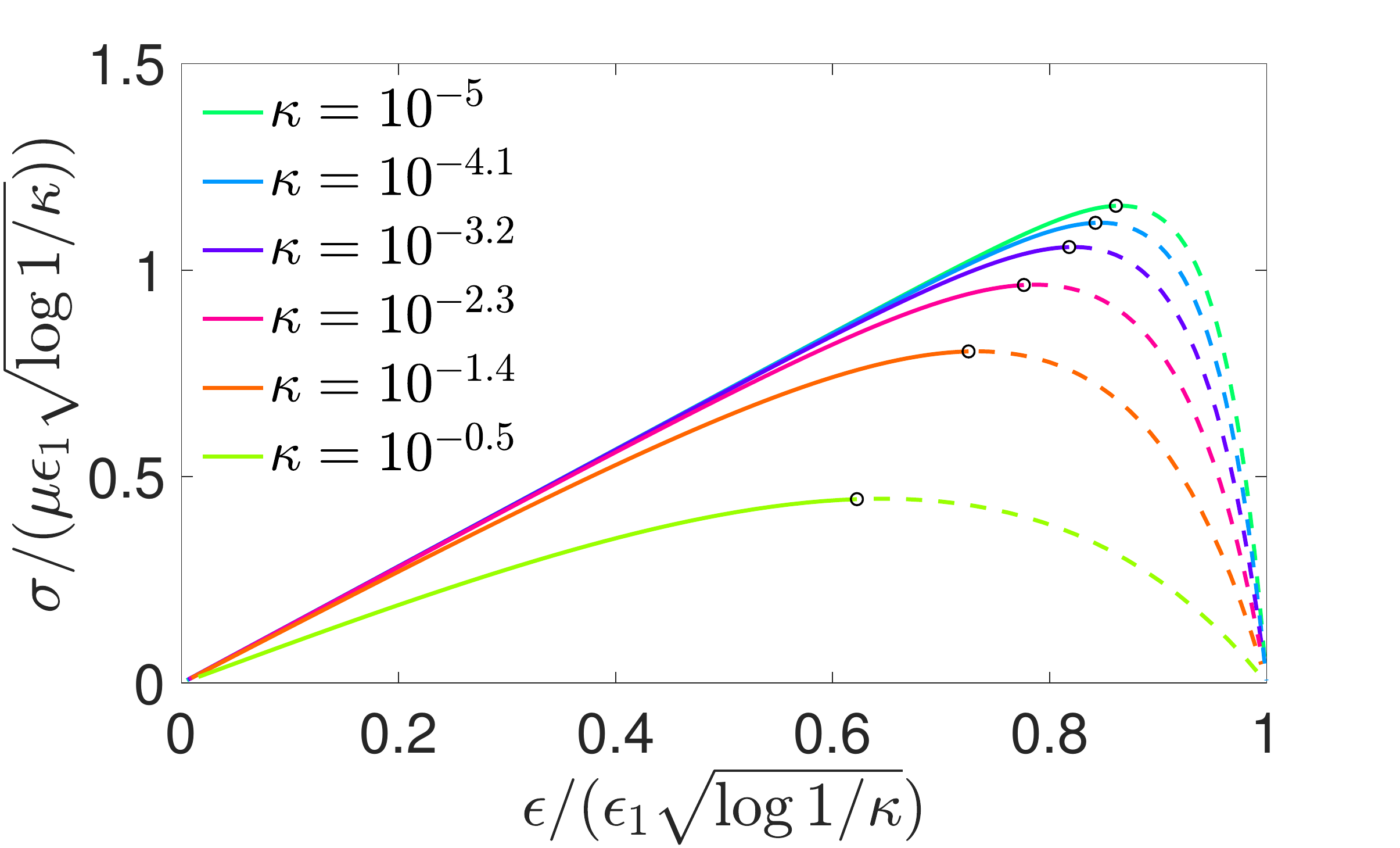}
\caption{ Stress {\it vs} strain curves for various values of $\kappa$ from $10^{-5}$ (top) to $10^{-1/2}$ (bottom) in the independent defect theory. { The curves are dashed beyond the local maximum, which we interpret as a yield stress.}
}\label{fig2}
\end{figure}

{\bf Renormalization of Elastic Quadrupoles: }
For quadrupoles as we consider here, the couplings $h, k_1, \gamma_{11}$ and $\gamma_{12}$ play no role. 
 We have
\eq{ \label{Ztau}
Z_\tau = \frac{\pi (a/\tau_c)^2 }{2 \beta k_2} e^{\frac{\beta a^2 j^2 \sigmabar^2}{k_2}} 
}
where $\sigmabar^2 = \half \sigmasbar_{ij} \sigmasbar_{ij}$. This fixes the dipole-moment scales and fluctuations defined as
\eq{ \label{fluc1}
\langle \tau_{ij} \rangle_\tau & = \half \tau_1 \delta_{ij} + \ffrac{1}{4} a_\tau \sigmasbar_{ij}/\sigmabar \\
\langle \tau_{ii}^2 \rangle_\tau & = \half \tau_2 (1+\chi) \label{fluc2} \\
\langle \tau_{ij} \tau_{ij} \rangle_\tau & = \ffrac{1}{4} \tau_2 (1+3\chi)
}
where $\langle \cdot \rangle_\tau$ is an expectation over a single defect:
\eq{
\langle \mathcal{A}(\tauh) \rangle_\tau \equiv Z_\tau^{-1} \int_{\tau} e^{-\beta V(\tauh)} \mathcal{A}(\tauh)
}
For quadrupoles $\tau_1=0$, $\chi = -1$, and 
\eq{
a_\tau & = -2\sqrt{2} j a^2\sigmabar/k_2 \\
\tau_2 & = -2 a^2 (k_2+ \beta a^2 j^2 \sigmabar^2)/(\beta k_2^2).
}
Consider first the non-interacting limit. The total free energy per unit area is
\eq{
\frac{1}{\Omega} F = \half \sigmabbar : \epsbbar - \frac{Z_\tau}{\beta a^2}
}
where $\epsbbar = \ffrac{1}{2\mu} \left[ \sigmabbar - \frac{\nu}{1+\nu} \delb \; \tr(\sigmabbar) \right]$ is a strain tensor. The first term in $F$ comes from the contribution of $\sigmabbar$ to the elastic energy. In the absence of defects, this is the only component of stress with a nonzero expectation value; hence $\sigmabbar$ is the constant component of elastic stress, and $\langle \sigmab \rangle - \sigmabbar$ is then the plastic stress. The latter can be computed from
\eq{
\frac{1}{\Omega} \frac{\p F}{\p \sigmabbar} = \frac{1}{2\mu} \left\langle \sigmab - \frac{\nu}{1+\nu} \delb \; \tr(\sigmab) \right\rangle
} 
leading to
\eq{
\sigma^2 \equiv \half \langle \sigmas_{ij} \rangle \langle \sigmas_{ij} \rangle = 2 \sigmabar^2 \left(1 - \frac{4 \mu j^2 Z_\tau}{k_2} \right)^2
}
This gives the total shear stress as a function of the elastic shear strain $\overline{\epsilon} = \sigmabar/\mu$. Introducing the strain scale 
\eq{ \label{eps1}
\epsilon_1 = (\beta \mu^2 a^2 j^2/k_2)^{-1/2}
}
this relation is plotted in Fig.\ref{fig2} for various values of
\eq{
\kappa = \frac{2\pi a^2 j^2 \mu}{\beta \tau_c^2 k_2^2}
}
from $10^{-5}$ (top) to $10^{-1/2}$ (bottom). Evidently once the strain is large enough, the stress begins to decrease with strain; we interpret the local maximum as a yield stress, { and consider the theory to only be reliable for smaller strain}. This phenomenon occurs because as strain is increased, more quadrupoles are excited, and each quadrupole counters the applied stress. It is a finite-temperature analog of the yielding scenario discussed in \cite{Dasgupta12,Dasgupta13a}. Quantitatively, the yield strain scales as $\epsilon_1$, with a logarithmic correction from $\kappa$:
\eq{ \label{epsy}
\epsilon_y^2 \approx \epsilon_1^2 \log(\epsilon_1^2/(\kappa\epsilon_y^2)) \sim \epsilon_1^2 \log(1/\kappa) \qquad \kappa \ll 1
} 
The sharpness of the transition is controlled by $\kappa$. Comparing with the expression for $Z_\tau$ (Eq.\eqref{Ztau}), we see that $\kappa$ is proportional to the defect density at zero strain. As this density increases, the transition becomes more smoothed out. This agrees with findings in \cite{Ozawa18,Popovic18}.

Let us now see how interactions complicate this picture. First, we notice that for quadrupoles the $k_2$ evolution equation reduces to $\p_t k_2 = -\ffrac{5\pi}{2} A \gamma_{22}^2$, which implies that $4k_2-5 \gamma_{22}$ is constant. Introduce the important constant
\eq{ \label{X}
X = 4 k_2^0 - 5 \gamma_{22}^0
}
Using \eqref{init1} we have $X = 4 (k_2^0 - 10 a_2/\mu)$. We choose units with bare values $a=\mu=1$ and fix $\nu=2/5$. The results then depend on $X$, on the temperature $\beta^{-1}$, the shear stress $\sigmabar$, and the dipole-moment scale $\tau_c$. Since $\tau_c$ sets the fugacity scale, it controls the defect density, and should be considered as a parameter set by the quench. At a qualitative level, lowering the temperature is similar to increasing $\tau_c$. 

\begin{figure}[t!]
\includegraphics[width=0.9\columnwidth]{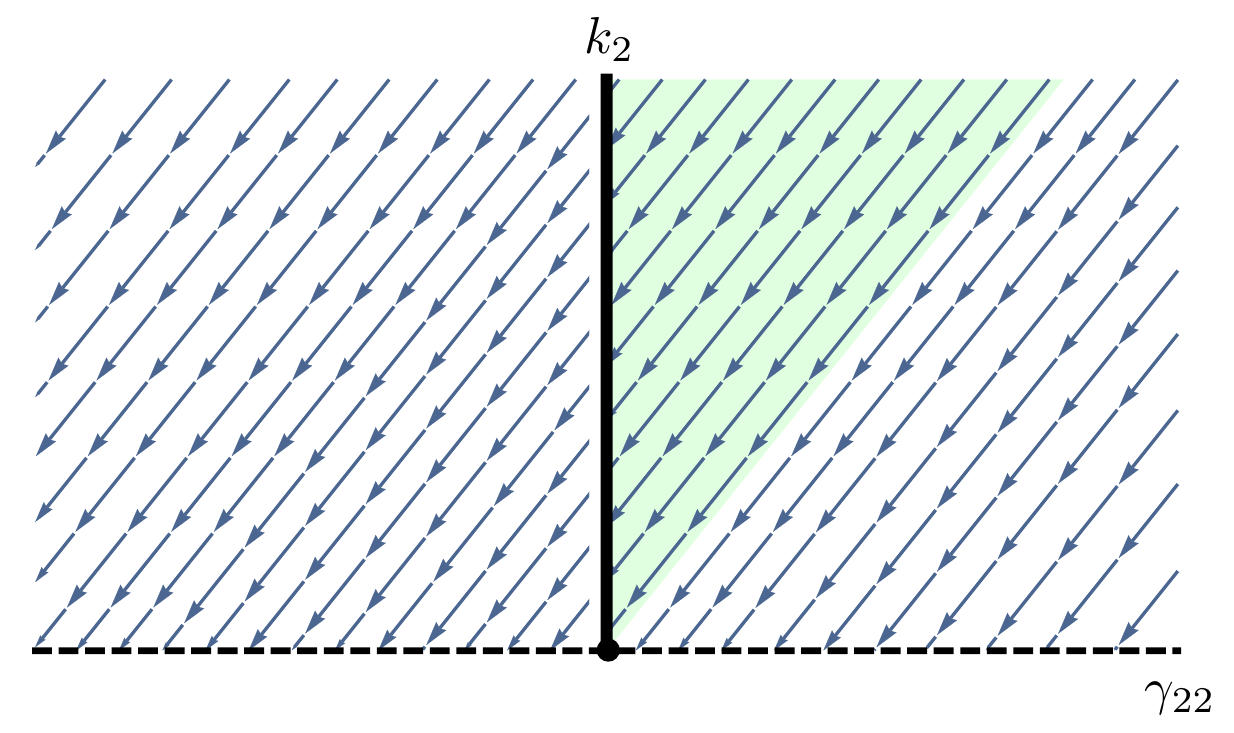}
\caption{ Projection of renormalization group flow onto $(\gamma_{22},k_2)$ space, or equivalently, complete flow for $\sigmabar=0$. There is a line of independent-defect fixed points along $\gamma_{22}=0$. Only trajectories beginning in the shaded region ($X>0$ and $\gamma_{22}>0$) end there; otherwise trajectories tend to the line $k_2=0$ where fluctuations diverge. 
}\label{RGflow1}
\end{figure}

Consider first the case $\sigmabar=0$. The RG flow in the $(\gamma_{22},k_2)$ plane is shown in Fig. \ref{RGflow1}. For $X>0$ and $\gamma_{22}>0$ the flow tends towards $\gamma_{22}=0$, which is {\it a line} of independent-defect fixed points; we call this the stable phase. Otherwise, the flow ends at $k_2=0$. In the latter case the fugacity $Z_\tau$ diverges, hence this corresponds to a proliferation of quadrupoles. These regimes are separated by the critical line $X=0$ ending at the origin. 

\begin{figure}[t!]
\includegraphics[width=\columnwidth]{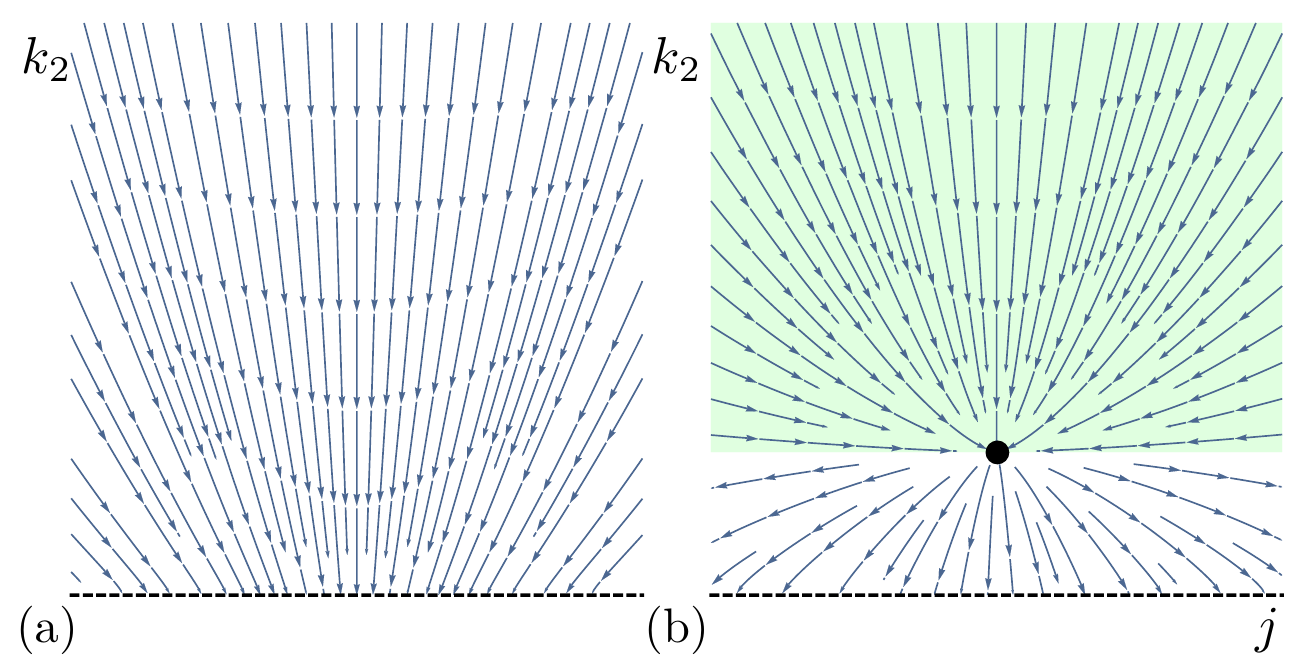}
\caption{ Renormalization group flow in $(j,k_2)$ space for (a) $X<0$ and (b) $X>0$. The basin of attraction of the fixed point manifold is shaded. Here $\beta=1,\tau_c=10,\sigma=0.1$ and the shown region is $|j|<1,0<k_2<1$.
}\label{RGflow2}
\end{figure}

When $\sigmabar>0$, the same picture is obtained (Fig. \ref{RGflow2}). In the stable phase, $j \to 0$, while in the unstable phase, $j$ tends to a non-universal constant, depending on its initial value. The fixed point in $(j,k_2)$ space is the line of fixed points shown in Fig. \ref{RGflow1}. 

In practice, these asymptotic behaviors are not always reached, because the RG flow is very slow. Consider $\sigmabar=0$, for which
\eq{\label{g22eq1}
\p_t \gamma_{22} = - 20 C_2 \;\frac{(\gamma_{22})^2}{(5\gamma_{22}+X)^2}
}
with $C_2=2\pi^2 a^2/(5\beta \tau_c^2)$. This can be integrated to
\eq{ \label{g22eq2}
\frac{5}{4} (\gamma_{22}^0-\gamma_{22}) \left[1 + \frac{X^2}{25 \gamma_{22}\gamma_{22}^0} \right] - \frac{X}{2} \log\frac{\gamma_{22}}{\gamma_{22}^0} = C_2 t,
}
where we recall that $t=\log a$. There are three regimes: for large enough $\gamma_{22}$, $\gamma_{22}^0-\gamma_{22} \lesssim 4|X|/5$, we have $\gamma_{22}^0-\gamma_{22} \sim \log a$, independent of the sign of $X$. Instead for smaller $\gamma_{22}$, this sign matters: for $X<0$ $\gamma_{22}$ tends to a constant at a finite scale, while for $X>0$ we find first a nontrivial power-law $\gamma_{22}/\gamma_{22}^0 \approx a^{-2C_2/X}$, and eventually $\gamma_{22} \sim 1/\log t$, these latter regimes being well-separated only for very small $|X|$. 

These results imply that there is a length scale $\xi$ below which all couplings evolve only logarithmically, and the system is stable. Above this length, either the system remains stable ($X>0$) and the couplings can show nontrivial power-law behavior, or the system is ultimately unstable ($X<0$) and fluctuations diverge. The critical length can be obtained by setting $k_2=0$ in \eqref{g22eq2}, for $X<0$. We find
\eq{ \label{xi1}
\xi \approx a \; e^{k_2^0/C_2} 
}
to leading order in $X$; this result then is also valid for $X>0$ to leading order. Since $\xi$ is exponential in the parameters, it can be astronomically large, in which case only the (transient) stable regime would be seen. In particular, for a system of linear size $L$, if $\xi > L$, then only the transient regime will be seen, and neither will the fixed-point be reached, nor will fluctuations diverge. For $\xi < L$, however, these two regimes will be distinguished.

\begin{figure}[t!]
\includegraphics[width=\columnwidth]{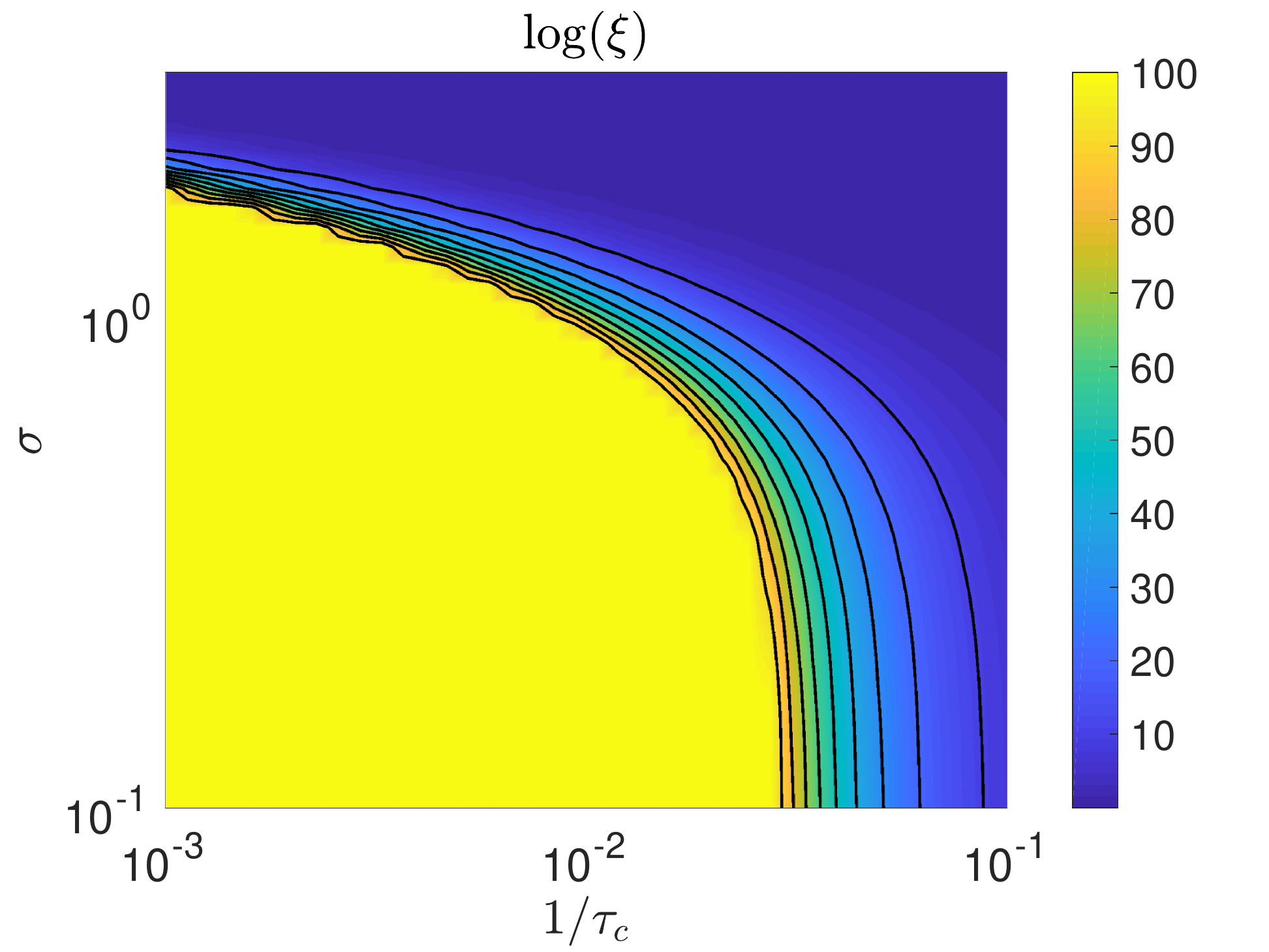}
\caption{ Contours of logarithm of length scale $\xi$ as a function of $\sigmabar$ and $1/\tau_c$, up to a maximum of $\log \xi = 100$. The length $\xi$ is exponentially large in the region closest to the origin. 
}\label{fig4}
\end{figure}

\begin{figure}[t!]
\includegraphics[width=\columnwidth]{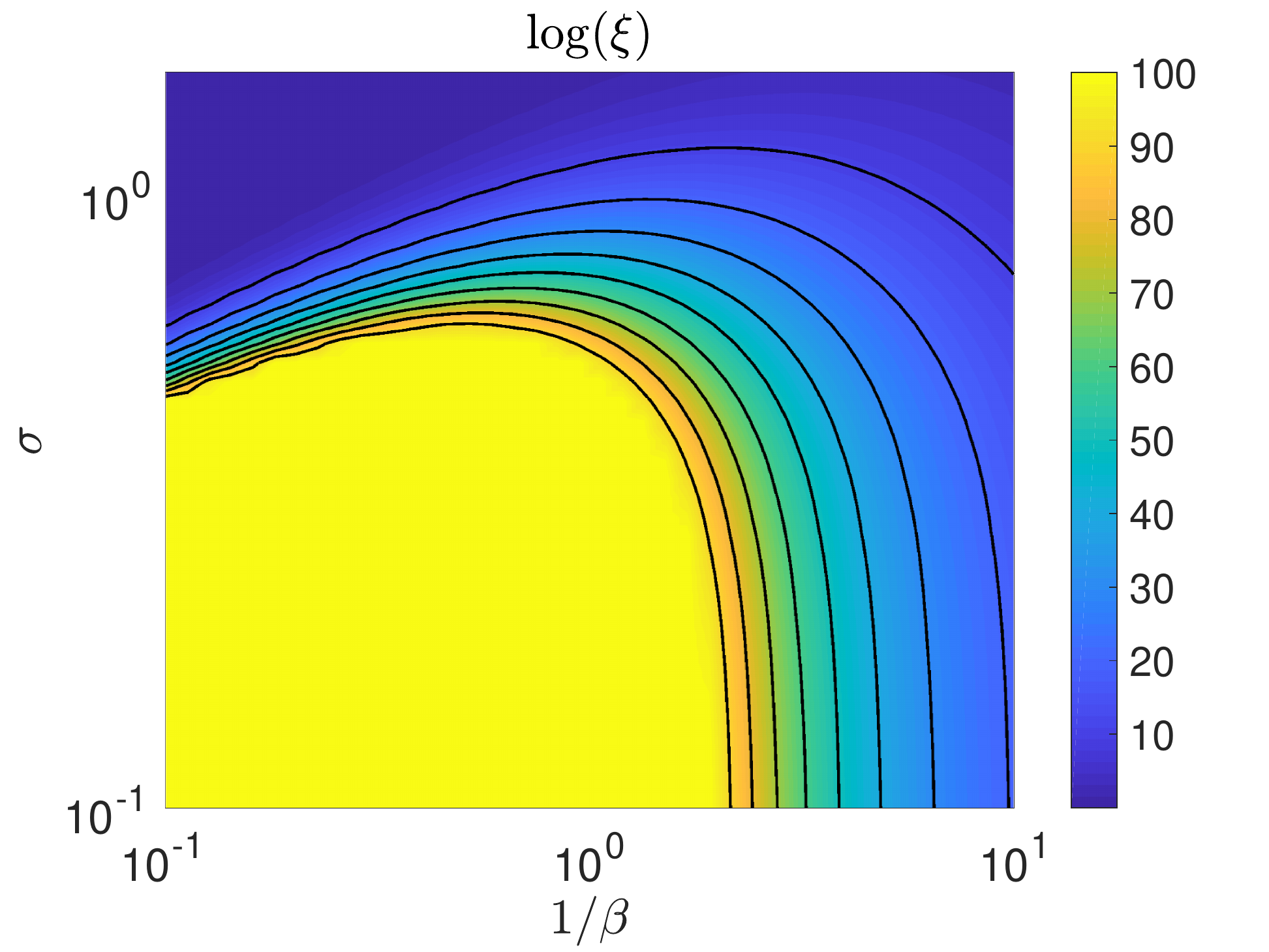}
\caption{ Contours of logarithm of length scale $\xi$ as a function of $\sigmabar$ and $1/\beta$, up to a maximum of $\log \xi = 100$. The length $\xi$ is exponentially large in the region closest to the origin. 
}\label{fig5}
\end{figure}

When $\sigmabar>0$, the phenomenology is similar. First we consider the unstable regime $X<0$. Again there is a length $\xi$ such that $k_2 \to 0$ at larger scales. In Fig. \ref{fig4} we show contours of $\log \xi$ as a function of $\sigmabar$ and $1/\tau_c$, up to a maximum of $\xi = e^{100}$ (Here $\beta \mu a^2=1$.). We can also study the transition at fixed $\tau_c$ while $\beta$ varies. In Fig. \ref{fig5} we show contours of $\log \xi$ at fixed $\tau_c = 50$ and varying $\beta$. At small enough temperatures and small enough stress, the length $\xi$ is exponentially large, so again the transition is avoided. 

When $X>0$, then $\xi$ is still relevant: below this scale, the RG flow is logarithmic, while above, there is a regime of power-law behavior before the fixed point is approached. To see this, we note that when $\beta a^2 j^2 \sigmabar^2/k_2 \ll 1$, we have $\gamma_{22}/\gamma_{22}^0 = (j/j_0)^{\sqrt{2}}$. Anomalous behavior of $\gamma_{22}$ thus carries over to $j$. 

The corrections to $\xi$ at finite $\sigmabar$ can be obtained from the RG equations. For simplicity we set $X=0$ and neglect the flow of $j$. Then 
%
\eq{
\log \xi/a & \approx \frac{g}{C_2} \int_0^{k^0_2/g} \frac{dk \; k}{k+1} e^{-1/k} \notag \\
& \approx \frac{k^0_2}{C_2} \left(1 - \frac{2 \overline{\epsilon}^2}{\epsilon_1^2} + \ldots \right) \qquad \overline{\epsilon} \ll \epsilon_1, \label{xi2}
}
where $g=\beta a^2 j^2 \sigmabar^2$. For $\overline{\epsilon} \ll \epsilon_1$ this reduces to \eqref{xi1} as expected. We can obtain a yield strain by finding when $\xi = L$. This leads to
\eq{ \label{epsy2}
\epsilon_y^2 \approx \frac{\epsilon_1^2}{2 \log 2}  \left[ 1 - \frac{C_2}{k_2^0} \log L/a + \ldots \right]
}
Comparing with \eqref{epsy} we see that interactions between defects have shifted the yield strain to a smaller value. This is consistent with expectations from the simple renormalization argument.

Note that all of these results rest on the weak-fugacity assumption $Z_\tau \ll 1$. This implies $\overline{\epsilon} < \epsilon_1$ and, at $\sigmabar=0$, $\beta \gg a^2/(k_2 \tau_c^2)$. 

To summarize this section, we find that there is a critical length $\xi$ such that couplings are only weakly scale-dependent on smaller scales. If the system scale $L<\xi$, then the solid is stable. When $L>\xi$, the behavior depends on the relative strength of the bare self-energy and interactions, represented by $X$ \eqref{X}. When $X>0$, corresponding to a large self-energy, the solid flows towards a fixed point with non-interacting defects. At scales larger than $\xi$, there is a regime in which couplings have anomalous power-law behavior, although this is only predicted near the limits of validity of the theory, i.e. when $k_2$ is small. Instead when $X<0$, corresponding to a moderate self-energy, the solid is ultimately unstable and $k_2 \to 0$. This corresponds to a divergence of fluctuations, and we interpret this spinodal transition as a yielding or melting transition, depending on the control parameter. These transitions are smoothly related, as shown in Figs. \ref{fig4},\ref{fig5}. 

However, since the weak-fugacity assumption of the RG calculation breaks down as the transition is approached, this transition may in fact disappear in a more complete theory, and in particular we cannot reliably extract information near the predicted transition. To confirm and extend the above results, we therefore proceed to a field-theoretic formulation, in which the defects are identically summed over. 

{\bf Dual field theory}: Following standard techniques, a defect model can be transformed into a dual field theory with a complex interaction, often with remarkable integrability properties. For example, the XY model maps onto the integrable sine-Gordon field theory, and the vector Coulomb gas to an extension thereof \cite{Chaikin00,Zhai19}. In this section we derive the novel field theory corresponding to \eqref{U2}, \eqref{V2}. 

The interaction $\tau_{ij}^a \tau_{kl}^b C_{ijkl}(\rv_{ab})/r_{ab}^2$ can be written as
\eq{ \label{I1}
I_{ab} = \tau_{ij}^a \tau_{kl}^b \frac{C_{ijkl}(\rv_{ab})}{r_{ab}^2} = \frac{1}{r_{ab}^2} \vec{v}^a \cdot \hat{E} \cdot \vec{v}^b
}
where $\vec{v}^a = \tau^a (1,\cos(2\theta_a),\sin(2\theta_a))$ and 
\eq{ \label{E}
\hat{E} = \begin{pmatrix} A & \alpha \cos(2\phi) & \alpha \sin(2\phi) \\
\tilde\alpha \cos(2\phi) & B + \gamma \cos(4\phi) & \gamma \sin(4\phi) \\
\tilde\alpha \sin(2\phi) & \gamma \sin(4\phi) & B - \gamma \cos(4\phi) \end{pmatrix},
}
where $\phi$ is the polar angle of $\vec{r}_{ab}$ and 
\eq{ \label{couplingvector}
A & = \gamma_{11} + \lambda_{11} + \ffrac{1}{4} \lambda_{22} \\
B & =\lambda_{11}+\ffrac{1}{2}\gamma_{22} \\
\tilde \alpha & = 2\lambda_{12} \\
\alpha & =2 \gamma_{12}+ 2\lambda_{12} \\
\gamma & =\lambda_{22} + \ffrac{1}{2} \gamma_{22}
}
Matrices of the form \eqref{E} belong to a matrix algebra, as shown in \cite{Lemaitre14}. 
Separating the interaction into the contributions from 
dilatant $(+)$ and compressive $(-)$ dipoles, we have
\eq{
& \sum_{a,b = 1, a\neq b}^{n'} I_{ab} = \sum_{a,b = 1, a\neq b}^{n} \frac{1}{r_{ab}^2} \sum_{\pm} \sum_{\pm'} \vec{v}_\pm^a \cdot \hat{E}(\rv_{ab}) \cdot \vec{v}_{\pm'}^b \notag \\
& \quad = \sum_{a,b = 1, a\neq b}^{n} \frac{1}{r_{ab}^2} \vec{v}_+^a \cdot \left[\hat{E}(\rv_{ab})+\hat{E}|_{\substack{\alpha \to -\alpha \\ \tilde\alpha \to -\tilde\alpha}}(\rv_{ab}) \right] \cdot \vec{v}_{+}^b
}
where in the last step we assume that dipoles are strictly bound into quadrupoles, as above\footnote{In a more general theory, we could consider a system composed of independent dilatant and compressive dipoles, with overall neutrality.}. For the initial values of the couplings, $A=0$, and the $\alpha$ and $\tilde \alpha$ terms vanish from neutrality, thus for the quadrupolar system only the lower-right 2x2 block of $\hat{E}$ is relevant, namely
\eq{
\hat{F}(\rv) = \frac{1}{r^2} \begin{pmatrix} B + \gamma \cos(4\phi) & \gamma \sin(4\phi) \\
\gamma \sin(4\phi) & B - \gamma \cos(4\phi) \end{pmatrix}
}
In Fourier space $\hat{F}$ is 
\eq{
\hat{F}(\qv) = -\pi \begin{pmatrix} b(q) + \gamma \cos(4\theta) & \gamma \sin(4\theta) \\
\gamma \sin(4\theta) & b(q) - \gamma \cos(4\theta) \end{pmatrix}
}
with $b(q) = 2B \log(q/\Lambda) < 0$ and $\theta$ the polar angle of $\qv$. 

For $B>0$ as we find, $\hat{F}(\qv)$ is positive-definite for $-b(q)>|\gamma|$. This inequality is violated in the UV regime, since $b(\Lambda)=0$. We will need to split the interaction into separate positive-definite and negative-definite parts, corresponding to stabilizing and destabilizing interactions, respectively.
To this end we consider an augmented operator $\tilde{\hat{F}}(\qv) = \hat{F}(\qv) + \eta k_2 \delb$, where $\eta$ is a parameter to be chosen such that $\tilde F$ is positive-definite at all $q$. This requires that
\eq{ \label{etaineq}
\eta > \frac{\pi|\gamma|}{k_2}.
}
Introducing the density
\eq{
\vec{w}(\rv) = \sum_a \tau^a (\cos(2\theta_a),\sin(2\theta_a)) \delta(\rv - \rv_a)
}
the dual field theory is derived by a standard method, explained in Appendix 3. It is given in terms of two vector fields $\zetav$ and $\epsv$, with a physical interpretation as plastic strain; they appear as a vector because we have introduced a Voigt-like representation of the tensorial interaction in \eqref{I1} above. $\epsv$ admits a straightforward interpretation: it has the same statistics as $-\epsv_0 + \eta k_2 (\wv-\wv_0)$, where $\vec{w}_0$ and $\epsv_0$ are constant background defect densities, to be fixed momentarily. $\zetav$, instead, couples to an imaginary field and is less transparent. 
The field theory has the nonlocal action
\eq{ \label{S}
S & = \half \beta \int_r \int_{r'} \zetav(\rv) \cdot \tilde{\hat{F}}^{-1}(\rv-\rvp) \cdot \zetav(\rvp) \\
& \qquad + \beta \int_r \left[ \frac{\epsilon(\rv)^2 + 2 \epsv(\rv) \cdot \epsv_0}{2\eta k_2} 
+ i\vec{w}_0 \cdot \zetav(\rv) - \frac{Z_0}{\beta a^2} e^{c \xi(\rv)^2} \right] \notag
}
where $\epsv_0 = B \log(R/a) \vec{w}_0 + \eta k_2 \wv_0 + j \vec{\sigma}$, $\vec{\xi}(\rv) =  i\zetav(\rv) + \epsv(\rv)$, $Z_0 = (\pi a^2)/(2\tau_c^2 \beta k_2)$, $c= \beta a^2/(2 k_2)$, and $\vec{\sigma} = \sigmabar (\cos 2\theta_\sigma, \sin 2\theta_\sigma)$. { The appearance of $\log(R/a)$ in $\epsv_0$ is due to the infrared divergence of $\hat{F}$ when acting on constants: $\int d^2r' \hat{F}(\rv-\rvp) \cdot \vec{w}_0 = B \log(R/a) \vec{w}_0$, for an asymptotically large domain of radius $R$.} It is implicitly assumed that any steric constraints on the defects are captured by a Debye cutoff $\Lambda = 2\pi/a$ in the field theory. Note that the form of the on-site potential $e^{c\xi^2}$ directly results from the Boltzmann measure on the defects. 



The background density $\wv_0$ is chosen such that $\zetav=0$ is a solution to the classical equation. This leads to $\epsv_c = -\epsv_0 + \eta k_2 \wv_0$, and 
when $\vec{\sigma}=0$ we find $\wv_0=0$ while for $\vec{\sigma}\neq 0$ we have $\wv_0 = -\rho j \vec{\sigma} /(B \log R/a)$ with
\eq{
\rho = \frac{Z_0 B \log R/a}{k_2} e^{c j^2 \sigmabar^2 (\rho-1)^2} \big(1-\rho\big)
}
This has a unique solution which for large systems is $\rho = 1 - k_2/(Z_0 B \log R/a) + \OO(1/\log^2 R/a)$.
The classical value for the partition function is then
\eq{
Z_C & = Z'_c \exp \left( \frac{Z_0 \Omega}{a^2} e^{c j^2 \sigmabar^2 (\rho-1)^2} \right) \\
& = \exp \left( \Omega \frac{Z_0}{a^2} + \OO(\Omega/\log R) + \ldots \right) \notag
}
and the Hessian is
\eq{
\hat{\mathcal{H}}(\rv) = \begin{pmatrix} \tilde{\hat{F}}^{-1}(\rv) & 0 \\ 0 & \hat{\delta F}^{-1} \end{pmatrix} + C \left[ \ell \vec{\sigma} \vec{\sigma} + c \delb \right] \begin{pmatrix} 1 & -i \\ -i & -1 \end{pmatrix} \delta(\rv) 
}
where $\hat{\delta F}(\rv) = \delta(\rv) \eta k_2 \delb$, $\ell = 2 j^2 (\rho-1)^2 c^2$ and $C = 2Z_0/(\beta a^2 (1-\eta)) e^{c j^2 \sigmabar^2 (\rho-1)^2}$.

To one-loop order we have \cite{Zinn-Justin96}
\eq{
Z_1 &= Z_C |\tilde F|^{-1/2} |\delta F|^{-1/2} |\mathcal{H}|^{-1/2} \notag \\
& = Z_C \exp \left( -\half \mbox{Tr} \log \left( \begin{pmatrix} \tilde F & 0 \\ 0 & \delta F \end{pmatrix} \mathcal{H} \right) \right) \\
& \propto Z_C \exp \left( -\half \Omega \int \frac{d^2 q}{(2\pi)^2} \log W(\qv) \right)
}
for an operator $W(\qv)$. For simplicity we consider only $\sigmabar=0$. Then after some work
\eq{
W(\qv) & = (1-\pi c C  b(q))^2 - (\pi c \gamma C)^2
}
Stability requires that $\log W(\qv)$ is always real and has at most integrable singularities. The most dangerous wavenumber is $q=\Lambda$ for which we require 
\eq{
k_2 > \pi \gamma Z_0 = \frac{\pi^2 a^2 \gamma}{2 \tau_c^2 \beta k_2}
}
Consistent with the RG analysis, the field theory breaks down when the self-energy is too small. Comparing with \eqref{xi1} we see that this can be written $k_2/\gamma_{22} > 5/(2 \log \xi/a)$. This should be compared with the condition to be in the stable phase determined above, i.e. $X>0$, which can be written $k_2/\gamma_{22} > 5/4$. The latter condition determined from the RG analysis is more stringent, apparently reflecting the logarithmic enhancement of fluctuations under renormalization.

Of course, this condition only reflects the stability to small fluctuations. The action has a term $e^{c\xi^2} = e^{-c\zeta^2} e^{+c\epsilon^2} e^{2i c\zetav \cdot \epsv}$, which contains potentially dangerous large fluctuations from $\epsilon$. It is shown in Appendix 3 that when $k_2$ is large enough, the $\epsilon$ field can be eliminated and the stability of the theory guaranteed (although the field remains imaginary). This holds for $k_2/\gamma > \pi$, which corresponds to the condition $X>0$, up to a numerical factor.


We thus find a well-defined dual field theory when $k_2 > \pi\gamma Z_0$, which is definitely stable for $k_2> \pi \gamma$. When $\sigmabar \neq 0$ it has an {\it imaginary} field, thus not strictly behaving as a statistical field theory.

Finally, since the operator $\hat{F}$ is local in Fourier space, the spectrum of $\mathcal{H}$ can be explicitly determined. The eigenvectors $\zeta_n(\rv)$ are anisotropic and quasi-localized and decay as $1/r^2$.  

\newcommand{\zetabar}{\overline{\zeta}}

{\bf Athermal and out-of-equilibrium systems: } The scenario described above holds when the stress tensor is sampled by a Boltzmann measure with Hamiltonian \eqref{H1}. 
{ One hypothetical experimental realization of this is a defected crystal in which all dislocations are strictly bound in pairs with equal and opposite Burger's vectors, in the special case where the separation $s$ between dislocations is fixed. From the KTHNY Hamiltonian it is straightforward to compute the interaction between two such pairs, which at distances $r \gg s$ takes the form of \eqref{H2}, as a function of the Burger's vectors and the separation vectors. 

We are not aware of any such crystal. However, by a re-interpretation of the theory, we expect the above scenario to hold for glasses, out of equilibrium. }
 Indeed, in this case measurements still occur in some ensemble specified by boundary conditions and experimental protocol. If stress is controlled, this is the stress ensemble \cite{Henkes09a,DeGiuli18,DeGiuli18a}, which in its field-theoretic version was argued to require only terms to Gaussian order \cite{DeGiuli18,DeGiuli18a}, in the generic case. For isotropic materials, one then finds an effective action exactly of the form $\beta H$ with $H$ as in \eqref{H1}, but with {\it effective} couplings with no {\it a priori} relationship to elastic moduli or temperature.

For example, such an ensemble can be derived by considering harmonic vibrations around an arbitrary inherent state, specified by its stress field\footnote{See the Supplementary Material in \cite{DeGiuli18}.}. After integrating out the phonons, the vibrational entropy of the state gives again \eqref{H1}, with couplings $\frac{\beta}{2\mu} =g, \frac{-\beta\nu}{2(1+\nu)\mu} = \eta$ in the notation of \cite{DeGiuli18}. The coefficients $g$ and $\eta$ both behave as $1/(D^d\mu^2)$ in $d$ dimensions, where $D$ is a length scale needed to regularize the measure; it has a natural interpretation as the particle size, up to $\OO(1)$ constants. Equivalently, we can consider the effective inverse temperature as $\beta \propto 1/(D^d\mu)$. If the system has only repulsive forces, then $\eta$ will be renormalized to $\eta \propto 1/(D^d \pbar^2)$ \cite{DeGiuli18a}. 

A natural hypothesis for the measure for glasses is then to combine contributions from the vibrational entropy and from the energy at the glass transition temperature. Regardless of these speculations, the hypothesis of an effective temperature has been explicitly tested in previous work on glasses \cite{Berthier00,Berthier02,Haxton07} \footnote{For applications to granular matter, see the discussion in \cite{DeGiuli18a}}. It was found that an effective temperature $T_{\eff}$ indeed controls the behavior under shear, and in particular that at any true temperature $T$, $T_{\eff}$ tends to a constant at vanishing strain rate. This is consistent with $T_{\eff}$ being tied to a shear modulus, as we suggest.  



{\bf Discussion: }

It has been argued on general grounds that stress fluctuations in amorphous solids are governed by a distribution of Boltzmann type, with effective parameters \cite{Henkes09a,DeGiuli18,DeGiuli18a}. Standard renormalization arguments imply that in its field-theoretic formulation, only terms to quadratic order are needed; the effective Hamiltonian is then equivalent to \eqref{H1}. Predictions for long-range stress correlations naturally followed from this theory, in excellent agreement with available data. This, however, presents a puzzle: since the theory is Gaussian in the gauge field, it is at bottom a non-interacting theory. But real amorphous solids yield under sufficient applied stress, and they can also liquify under heating. A non-interacting theory cannot support any phase transitions. What then is the missing ingredient in the theory, which controls the onset of these transitions?

We have shown here that localized excitations can fulfill this role. Indeed, by including them in the field theory, a transition appears, which we interpret as yielding if induced by stress, or melting if induced by temperature. This transition is fundamentally one of strong coupling, hence the behavior very near the transition is inaccessible with the present perturbative method, and indeed may disappear in a more general treatment. However, renormalization group and field-theoretic calculations are in approximate agreement concerning its location. Moreover, since the onset of instability is so abrupt, the present theory is valid until close to the predicted transition. So long as the fugacity $Z_\tau$ is small, the defects are only a small perturbation to the free energy. We thus explain why the Gaussian theory correctly predicts stress correlations in such a large range of parameters, and yet will eventually break down at the yielding/melting transition.  


The renormalization group method sheds some light on the organization of states within the stable solid phase. For example, the RG shows that the characteristic dipole-moment scale $\tau$ varies with scale $a$ as $\tau \sim a$ in 2D, generalizing to $\tau \sim a^{d/2}$ in $d$ dimensions. This exponent defines a fractal dimension, reminiscent of the fractal dimension of plasticity avalanches in amorphous solids. The latter has been studied both in steady-state yielding, where it is close to $1$ in 2D and close to $3/2$ in 3D \cite{Lin14,Lin15a,Ferrero19,Tyukodi19}, and in the quasi-elastic regime at small strain, where it is significantly smaller \cite{Franz17,Shang19}. Since the present theory applies in the stable solid, why the prediction $d/2$ is close to, but distinct from, the steady-state yielding result remains to be clarified. Marginal stability, known to be present in many amorphous solids, is likely playing a role \cite{Muller14,Berthier19,Charbonneau14,Lin14,Lin14a,Lin15a,Lin16,Biroli16,Urbani17,Shimada18,Scalliet19,Ji19}. 
 This could be investigated by simulating the dynamics of the present model, while the quantum formalism may be useful for an analytical treatment \cite{Beekman17}.




We found that yielding is a transition of spinodal type. Recent works indeed present evidence that yielding is of this form \cite{Procaccia17,Parisi17}, but the tools employed in these works are agnostic regarding the microscopic mechanism. Consistent with \cite{Dasgupta12,Dasgupta13a}, we have shown that localized quadrupolar defects will generally renormalize to create large-scale instability. In our model, the defects are treated on the same footing as the phonons; their magnitude and orientation are dynamical. It is very important to see what happens when some properties of the defects are considered quenched \cite{Nandi16}, since disorder necessarily controls dynamics of plasticity. 

We found that the transition demarks a stable phase around the origin in $(\sigmabar,1/\tau_c,1/\beta)$ space (Figs 4,5). The transition varies continuously with these parameters, but is more abrupt when induced by stress than by temperature. As the zero-strain defect density increases, the yielding transition becomes smoothed out, as found in \cite{Ozawa18,Popovic18}. We also find that the abrupt transition can disappear entirely when the bare self-energy is large enough, consistent with the transition between brittle and ductile failure recently discussed in \cite{Ozawa18,Popovic18}.

{ More precisely, the renormalization group calculation indicates that there is a length scale $\xi$ beyond which the system can be unstable, if $\xi$ is smaller than the system size $L$. This scale depends exponentially on parameters (Eq.\eqref{xi2}) and can thus be astronomically large. This sensitive dependence on parameters corresponds precisely to a weak, logarithmic, system-size dependence of the yield strain (Eq.\eqref{epsy2}). Such a logarithmic dependence can be seen in Fig.8 of Ref. \cite{Procaccia17}. It is currently unclear whether $\xi$ is related to the correlation length of solid domains derived in \cite{Maier17}, which diverges proportional to the relaxation time as the glass transition is approached from the liquid side. } 

We have not discussed what happens in the yielded phase. Since each quadrupole can be considered a bound state of a dilatant and compressive dipole, and the transition corresponds to a proliferation of quadrupoles, { as yielding is approached it becomes entropically unfavourable for the dipoles to remain strictly bound in pairs; this opens the possibility for collective excitations involving multiple pairs of dipoles. Taken to the limit, this would give a neutral plasma of dilatant ($\tau>0$) and compressive ($\tau<0$) force dipoles. It is possible that this unbinding is related to the increasing avalanche size and softening of the pseudo-gap observed as yielding is approached in elasto-plastic models \cite{Lin15a}. }


We have focussed on amorphous solids deep in the jammed phase, where a continuum approach with defects is appropriate. It has been shown in \cite{Shimada18} that as the jamming transition is approached, the quasi-localized modes have a growing core, and become the anomalous modes associated to the jamming point \cite{DeGiuli14,Lerner14,Yan16}. This suggests that a full treatment of the unjamming point in the present framework may require a more sophisticated description of mode cores than is accounted for by self-energies.

Finally, it would be useful to extend this theory to three dimensions. An analog of the Airy stress function exists but in this case the theory has a {\it bona-fide} gauge freedom \cite{DeGiuli18a}. 


{\bf Acknowledgments: } I am grateful to G. Biroli, T. Sulejmanpasic, G. Tarjus, and F. Zamponi for conversations at an early stage of this work, to E. Lerner for frequent discussions, and to M. Wyart and F. Zamponi for insightful comments on the manuscript.

\bibliography{../Glasses} 
\vfill

\vfill
\begin{widetext}

{\bf Appendix 1. Derivation of the defect energy. } In a planar elastic continuum, the response in stress to a point force $\vec{f}$ at $\vec{r}_0$ can be written in terms of the Airy stress function $\psi(\rv)$. In a complex notation with $f = f_x + i f_y, z_0 = r_{0,x} + i r_{0,y}$ the result is (p. 268 in \cite{Sokolnikoff56})
\eq{
\psi(z; f, z_0) = \frac{-\overline{z} f}{4\pi(1+c)}  \log \delta z + \frac{c \overline{f}}{4\pi(1+c)} \left[ \delta z \log \delta z - \delta z \right] + \mbox{c.c},
}
where $\delta z = z-z_0, $ $c=(\lambda+3\mu)/(\lambda+\mu)$, and c.c means complex conjugate. We now consider a pair of forces $\pm \vec{f}$ applied to $\rv \mp \vec{s}$, where $\vec{s} \propto \vec{f}$, in order to preserve both force and torque balance. The response is
\eq{
D(z; f, z_0, s_0) = \psi(z; f, z_0-s) + \psi(z; -f, z_0+s)
}
with $s = s_{0,x}+i s_{0,y}$. Expanding this in $s$ we find
\eq{
D(z; f, z_0, s_0) = \frac{\tau c}{2\pi(1+c)} \log |z-z_0|^2 - \frac{\tau}{\pi(1+c)} \cos(2\theta-2\theta_0) + \OO\left(\frac{\tau |s|^2}{|z|^2}\right),
}
where $f = |f| e^{i \theta_0}$, $z = |z| e^{i \theta}$, and $\tau = \vec{f} \cdot \vec{s}$. The first term is monopolar while the second is quadrupolar. It is convenient to adopt the notation of \cite{Moshe15a}; see Table 1 therein. We write
\eq{
D(\rv; f, z_0, s_0) = - \psi_0(\rv-\rv_0; M) - \psi_2(\rv-\rv_0; \hat{Q})
}
where 
\eq{
\psi_0(\rv; M) = \frac{-Y M}{2\pi} \log |r|
}
and
\eq{
\psi_2(\rv; \hat{Q}) = \frac{Y}{16\pi} \hat{r}\hat{r} : \hat{Q}
}
are canonical monopolar and quadrupolar defects, respectively. Here $M$ is the scalar monopolar charge, 
\eq{
\hat{Q} = Q \begin{pmatrix} \cos(2\theta_0) & \sin(2\theta_0) \\ \sin(2\theta_0) & -\cos(2\theta_0) \end{pmatrix}
}
is the tensor quadrupolar charge, and $Y= 2(1+\nu)\mu$ is Young's modulus. We define an elastic energy functional as
\eq{
E[\sigma^1,\sigma^2] = \ffrac{1}{2Y} \int_r A_{ijkl} \sigma^1_{ij} \sigma^2_{kl}
}
with $A_{ijkl} = (1+\nu) \delta_{ik} \delta_{jl} - \nu \delta_{ij} \delta_{kl}$. The total stress field $\sigmab$ is decomposed into $\sigmab = \sigmabbar + \sigmab^P + \sum_a \sigmab^D_a$, where $\sigmabbar$ is a constant stress, $\sigmab^P$ are the transverse phonons, and $\sigmab^D_a$ is the $a^{th}$ defect. The latter two components are written in terms of Airy scalar fields using the double-curl operator $(\nabla \times \nabla \times)_{ij} = (\epsb \cdot \nabla)_i  (\epsb \cdot \nabla)_j =  \epsilon_{ik} \p_k \epsilon_{jl} \p_l$, i.e. $\sigmab^P = \nabla \times \nabla \times \psi$ and $\sigmab^D_a = \nabla \times \nabla \times D_a$. The energy is then
\eq{
E[\sigma,\sigma] = \overline{E} + E[\sigma^P,\sigma^P] + \sum_a E[\sigmab^D_a,\sigmab^D_a] + 2 E[\sigmabbar,\sigmab^P] + 2 \sum_a E[\sigmabbar,\sigmab^D_a] + 2 \sum_a E[\sigmab^P,\sigmab^D_a] + \sum_{a \neq b} E[\sigmab^D_a,\sigmab^D_b] 
}
We have $\overline{E} = \half \Omega \sigmabbar : \epsbbar$ where $\epsbbar = \ffrac{1}{2\mu} \left[ \sigmabbar - \frac{\nu}{1+\nu} \delb \; \tr(\sigmabbar) \right]$ is a strain tensor. The constant component of stress $\sigmabbar$ is orthogonal to the phonons and $E[\sigmabbar,\sigmab^P]=0$. Using results from Table 2 in \cite{Moshe15a} we have
\eq{
2 E[\sigmabbar,\sigmab^D_a] & = 2 M_a \tr \sigmabbar - \ffrac{1}{4} \hat{Q}_a : \sigmabbar
}
and
\eq{ \label{defectdefect}
2 E[\sigmab^D_a,\sigmab^D_b] = \frac{Y}{2\pi r_{ab}^2} \left[ M_a \Qb_b + M_b \Qb_a \right] : \rh_{ab} \rh_{ab} + \frac{Y}{16\pi r_{ab}^2} \Qb_a \Qb_b :: \left[ 2 \rh_{ab} \rh_{ab} \rh_{ab} \rh_{ab} - \rh_{ab} \;\delb \;\rh_{ab} \right],
}
where $\rv_{ab} = \rv_a - \rv_b$, $\rh_{ab}=\rv_{ab}/|\rv_{ab}|$, and we recall that we are using a notation in which all tensor contractions are explicitly indicated, e.g. $\Qb \Qb :: \rh \delb \rh = Q_{ij} Q_{kl} r_i \delta_{jk} r_l$. 


The interaction energy between a defect and another stress field $\sigmab^P = \nabla \times \nabla \times \psi$ can be written as $E = \half \int_r \psi K_G$ in terms of the defect curvature $K_G$, which for a monopole and quadrupole is $-2M \nabla^2 \delta(\rv-\rv_0)$ and $\ffrac{1}{4} \Qb : \nabla \nabla \delta(\rv-\rv_0)$, respectively \cite{Moshe15a}. The phonon-defect interaction energy is then
\eq{
2 E[\sigmab^P,\sigmab^D_a] & = 2\int_r \psi(\rv) \left[-2M_a \nabla^2 \delta(\rv-\rv_a) + \ffrac{1}{4} \Qb_a : \nabla \nabla \delta(\rv-\rv_a) \right] \\
& = \frac{2}{\Omega} \sum_{\qv} \psi_{\qv} \; e^{i \qv \cdot \rv_a} \left[2q^2 M_a  - \ffrac{1}{4} \qv \qv : \Qb_a \right] \\
& \equiv \sum_{\qv} \psi_{\qv} \;\qv \qv : \hat{A}^a_{\qv} \\
& = \sum_{\qv \in \text{BZ}_U} \left[ \psi_{\qv} \;\qv \qv : \hat{A}^a_{\qv} + \psi^\dagger_{\qv} \;\qv \qv : \hat{A}^a_{-\qv} \right]
}
defining a tensorial operator $\hat{A}$, and then writing the sum over BZ${}_U$, the half of the first Brillouin zone in the upper-half plane. 

The phonon self-interaction has the form
\eq{
E[\sigma^P,\sigma^P]  & = \frac{1}{2Y} \int_r \left[ (1+\nu) \left( (\epsb \cdot \nabla) (\epsb \cdot \nabla) \psi \right) : \left( (\epsb \cdot \nabla) (\epsb \cdot \nabla) \psi \right) - \nu \left( \nabla^2 \psi \right)^2 \right] \\
& = \frac{1}{2Y} \int_r \left[ (1+\nu) \left(\nabla \nabla \psi : \nabla \nabla \psi \right) - \nu \left( \nabla^2 \psi \right)^2 \right] \\
& = \frac{1}{2Y\Omega} \sum_{\qv} \left[ (1+\nu) \qv \qv \psi_{\qv} : \qv \qv \psi_{-\qv} - \nu q^4 \psi_{\qv} \psi_{-\qv} \right] \\
& = \frac{1}{2Y\Omega} \sum_{\qv} q^4 \psi_{\qv} \psi_{-\qv} \\
& = \frac{1}{Y\Omega} \sum_{\qv \in \text{BZ}_U} q^4 \psi_{\qv} \psi^\dagger_{\qv}
 }
In the interest of future work we will change $Y$ to $\tilde Y$ to distinguish it from the Young's modulus appearing elsewhere. We can now integrate out the phonons:
\eq{
\int \D\psi \; e^{-\beta E[\sigma^P,\sigma^P]} e^{- 2\beta \sum_a E[\sigma^P,\sigma^D_a]}
& = \prod_{\qv \in \text{BZ}_U} \int d^2\psi_{\qv} \; e^{-\frac{\beta}{\tilde Y \Omega} q^4 \psi_{\qv} \psi^\dagger_{\qv}} e^{-\beta \psi_{\qv} \;\qv \qv : \hat{A}^a_{\qv} - \beta \psi^\dagger_{\qv} \;\qv \qv : \hat{A}^a_{-\qv} } \\
& = \prod_{\qv \in \text{BZ}_U} \frac{\pi \tilde Y \Omega}{\beta q^4} \exp \left( \beta \tilde Y \Omega \;\qh \qh \qh \qh :: \sum_{a,b} \hat{A}^a_{\qv}\hat{A}^b_{-\qv} \right) \\
& \propto  \exp \left( \frac{2\beta \tilde Y}{\Omega} \; \sum_{a,b} \hat{K}_a \hat{K}_b :: \sum_{\qv} \qh \qh \qh \qh \;e^{i \qv \cdot \rv_{ab}} \right)
}
where $\hat{K}_a = 2 M_a \delb - \ffrac{1}{4} \Qb_a$. We need to evaluate
\eq{
I_{ijkl}(\rv) & = \frac{1}{\Omega} \sum_{\qv} \qh_i \qh_j \qh_k \qh_l \;e^{i \qv \cdot \rv} \\
& = I_1(r) \left[ \delta_{ij}\delta_{kl} + \delta_{ik}\delta_{lj}+\delta_{il}\delta_{jk} \right] + I_2(r) \rh_i \rh_j \rh_k \rh_l \notag \\
& \qquad + I_3(r) \left[ \rh_i \rh_j \delta_{kl} + \rh_i \rh_l \delta_{jk}  + \rh_i \rh_k \delta_{jl}  + \rh_j \rh_k \delta_{il} +\rh_j \rh_l \delta_{ik} + \rh_k \rh_l \delta_{ij} \right]
}
where the second form follows since $I_{ijkl}$ depends only on $\rv$, is symmetric in all indices, and is invariant under a spatial reflection. In $d$ dimensions we have
\eq{
I_{iikk} & = \frac{1}{\Omega} \sum_{\qv} e^{i \qv \cdot \rv} \qquad \quad = I_1 (d^2+2d) + I_2 + I_3 (2d+4)\\
\rh_i \rh_j I_{ijkk} & = \frac{1}{\Omega} \sum_{\qv} (\qh \cdot \rh)^2 e^{i \qv \cdot \rv} = I_1 (d+2) + I_2 + I_3 (d+5) \\
\rh_i \rh_j \rh_k \rh_l I_{ijkl} & = \frac{1}{\Omega} \sum_{\qv} (\qh \cdot \rh)^4 e^{i \qv \cdot \rv} = 3 I_1 + I_2 + 6 I_3
}
and one can compute 
\eq{
\hat{K}_a \hat{K}_b :: \hat{\hat{I}} & = -\half (M_a \Qb_b + M_b \Qb_a) : \rh\rh \left[ I_2+(d+4)I_3 \right] + \ffrac{1}{16} \Qb_a \Qb_b :: \left[ I_2 \rh \rh \rh \rh + 2 I_1 \hat{\hat{\Delta}} + 4 I_3 \rh \delb \rh \right] \\
& \qquad + 4 M_a M_b \left[ d(d+2) I_1  + I_2 + 2(d+2) I_3 \right]
}
where $\Delta_{ijkl} = \delta_{ik} \delta_{jl}$. We now restrict to $d=2$ with an isotropic Brillouin zone $0<q<\Lambda$, and evaluate the sums in the continuum. For $r=0$ we have 
\eq{
I_1(\rv=0) = \Lambda^2/(32\pi), \quad I_2(\rv=0) = I_3(\rv=0)=0, 
}
while for $r \neq 0$ we have
\eq{
I_1(\rv \neq 0) & = \frac{1}{4\pi r^2} \left[ 1 -2(r \Lambda)^{-1} J_1(r\Lambda)  - \ffrac{2}{3} r \Lambda J_1(r\Lambda) \right] \\
I_2(\rv \neq 0) & = \frac{1}{4 \pi r^2} \left[ 8 + 16 J_0(r\Lambda) - 48 (r \Lambda)^{-1} J_1(r\Lambda)  \right] \\
I_3(\rv \neq 0) & = \frac{1}{4 \pi r^2} \left[ -2 - 2 J_0(r\Lambda) + 8(r \Lambda)^{-1} J_1(r\Lambda) + \ffrac{2}{3} r \Lambda J_1(r\Lambda) \right]
}
Introduce a spatial-coupling matrix
\eq{ \label{P1}
\quad \hat{P}(\rv) = \begin{bmatrix} \cos 2\phi & \;\;\sin 2\phi \\ \sin 2\phi & -\cos 2\phi \\ \end{bmatrix},
}
where $\phi$ is the polar angle of $\rv$. Then $\Qb : \rh \rh = \half \Qb : \Pb$. Altogether the effective Hamiltonian for the defects is
\eq{ \label{appH}
H & = \sum_a \left[ 2 M_a \tr \sigmabbar - \ffrac{1}{4} \hat{Q}_a : \sigmabbar + E[\sigmab^D_a,\sigmab^D_a] - 2 \tilde Y \frac{\Lambda^2}{32\pi}  \left[ \ffrac{1}{4} Q_a^2 + 32 M_a^2 \right] \right] \\ 
& \qquad + \sum_{a < b} \left[ 2\frac{Y}{2\pi r^2} \left[ M_a \Qb_b + M_b \Qb_a \right] : \half \Pb + 2 \frac{Y}{16\pi r^2} \Qb_a \Qb_b :: \left[ 2 \Pb \Pb - \rh \;\delb \;\rh \right]  - 4 \tilde Y \hat{K}_a \hat{K}_b :: \hat{\hat{I}} \right]
}
where $r=r_{ab}$ everywhere and $\Pb=\Pb(\rv_{ab})$ everywhere. We note that the phonons induce a destabilizing self-interaction between the defects. This must be compensated by the defect self-energy $E[\sigmab^D_a,\sigmab^D_a]$, which requires regularization. A naive estimation obtained by replacing $\delta(\rv)$ by $1/(\pi a^2)$, where $a$ is a core size, yields terms of the same form as the phonon-mediated self-interaction, but with the opposite, stabilizing, sign. Physically the resultant self-energy must be positive, but its magnitude then depends sensitively on $a \Lambda$, i.e. the core size compared to the UV cutoff for the phonons. Since this quantity is not well constrained, we simply lump together these terms into
\eq{
E_a^{\text{self}} = \ffrac{\mu^2}{2a^2} \left[ \ffrac{16}{c} (k_1+\half k_2) M_a^2 + \ffrac{1}{8} k_2 Q_a^2 \right],
}
where $a$ is a microscopic length scale and the $k_i$ have units of inverse shear modulus; this particular form is chosen for later convenience. We expect $k_i \propto 1/\mu$ with $\OO(1)$ coefficients.

\begin{figure}[t!]
\includegraphics[width=\columnwidth]{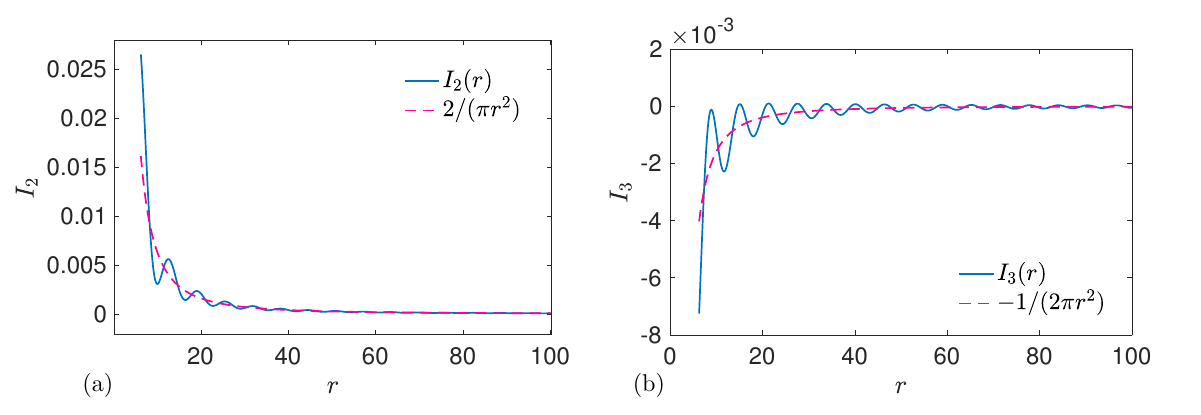}
\caption{ Comparison of functions $I_2(r)$ and $I_3(r)$ with simplified scale-free forms, $\tilde I_2(r) = 2/(\pi r^2)$ and $\tilde I_3(r) = -1/(2\pi r^2)$, respectively. Here we have taken $\Lambda=1$ so that the smallest $r$ is $2\pi$. 
 Note that $I_1$ is obtained from $I_1(r) = -I_3(r) - I_2(r)/8$ for $r>0$.
}\label{figapp}
\end{figure}


In the main text we adopt the notation of a tensorial dipole moment, 
\eq{ \label{apptau1}
\hat{\tau} = \half \tau \begin{bmatrix} 1+\cos 2\theta & \sin 2\theta \\ \sin 2\theta & 1-\cos 2\theta \\ \end{bmatrix} = \ffrac{2\mu}{c} M \delb + \ffrac{\mu}{4} \Qb,
}
and rewrite the interaction as a quadratic form in $\tauh$. 

Note that the functions $I_i$ have secular components $\propto 1/r^2$, and components that fluctuate on the scale $r \Lambda$. For simplicity, in this work we consider the scale-free form in which the fluctuating components are neglected; see Fig.\ref{figapp}. 
 Future work will critically examine this simplification.


{\bf Appendix 2: Derivation of renormalization group flow.} 

We consider two fixed charges, $\tauh$ and $\tauh'$, and see how their interaction, including their self-interaction, is renormalized by excited charges. The role of fugacity is played by the single-defect partition function $Z_\tau = \int_\tau e^{-\beta V(\tauh)}$. At leading order in $Z_\tau$, we just consider one excited charge. Let $\rvp = \rv - \vec{s}$ where $\vec{s}$ is the vector from $\tauh$ to $\tauh'$. We perform the computation where each charge has a hard core of radius $a$. The renormalized interaction between the external charges, including self-interactions, is
\eq{
& e^{-\beta\mathcal{V}(\tauh) -\beta\mathcal{V}(\tauh')} e^{-\beta\mathcal{C}_{ijkl} \tau_{ij} \tau'_{kl}/s^2} \notag \\ \label{RG1}
& \qquad = e^{-\beta V(\tauh) - \beta V(\tauh')} e^{-\beta C_{ijkl} \tau_{ij} \tau'_{kl}/s^2} \frac{\left[ 1 + Z_\tau \int_\Omega \frac{d^2 r}{a^2} \left\langle e^{-\beta C^r_{ijkl} \tau_{ij} \tau{''}_{kl}/r^2} e^{-\beta C^{r'}_{ijkl} \tau'_{ij} \tau{''}_{kl}/(r')^2} \right\rangle_{\tauh''} + \OO(Z_\tau^2) \right]}{1+Z_\tau \int_\Omega \frac{d^2 r}{a^2} \left\langle 1 \right\rangle_{\tauh''} + \OO(Z_\tau^2)}  
}
where the integration domain is $\Omega = \{r>a,r' > a\}$. The denominator ensures that the interaction is not renormalized when the excited charge does not interact with the fixed charges. 
Then
\eq{ \label{RG2}
\mathcal{V}(\tauh) +  \mathcal{V}(\tauh') + \frac{1}{s^2} \mathcal{C}_{ijkl} \tau_{ij} \tau'_{kl} = V(\tauh) + V(\tauh') + \frac{1}{s^2} C_{ijkl} \tau_{ij} \tau'_{kl} - \frac{Z_\tau}{\beta} \int_\Omega \frac{d^2 r}{a^2}\left\langle I(\rv,\rvp,\tauh'') \right\rangle_{\tauh''} + \OO(Z_\tau^2) 
}
where
\eq{
I(\rv,\rvp,\tauh'') & = e^{-\beta C^r_{ijkl} \tau_{ij} \tau{''}_{kl}/r^2} e^{-\beta C^{r'}_{ijkl} \tau'_{ij} \tau{''}_{kl}/(r')^2}-1 
}
where the superscript $r$ indicates that $\hat{P}$ depends on $\rh$, and where we suppress dependence on $\tauh$ and $\tauh'$.

Following \cite{Jose77,Nelson79}, we can organize \eqref{RG2} into a renormalization group transformation by splitting $\Omega$ into the shells $a < r < ba$ and $a < r' < ba$ and the remainder $\Omega(b) \equiv \{r> ba,r' > ba\}$. Multiplying through by $a^2/\tau_c^2$ we can write  
\eq{ \label{RG3}
\mathcal{\tilde V}(\tauh) +  \mathcal{\tilde V}(\tauh') + \frac{a^2}{s^2} \mathcal{C}_{ijkl} \frac{\tau_{ij} \tau'_{kl}}{\tau_c^2} & = \tilde V(\tauh) + \tilde V(\tauh') + \frac{a^2}{s^2} C_{ijkl} \frac{\tau_{ij} \tau'_{kl}}{\tau_c^2} - \frac{Z_\tau a^2}{\beta \tau_c^2} \int_a^{ba} \frac{r dr}{a^2} \int_{\rh} \left\langle I(\rv,\rvp,\tauh'') \right\rangle_{\tauh''} \notag \\
& \qquad - \frac{Z_\tau a^2}{\beta \tau_c^2} \int_a^{ba} \frac{r' dr'}{a^2} \int_{\hat{r'}} \left\langle I(\rv,\rvp,\tauh'') \right\rangle_{\tauh''} - \frac{Z_\tau a^2}{\beta \tau_c^2} \int_{\Omega(b)} \frac{d^2 r}{a^2}\left\langle I(\rv,\rvp,\tauh'') \right\rangle_{\tauh''}
}
where $\mathcal{\tilde V}=(a/\tau_c)^2 \mathcal{V}, \tilde V = (a/\tau_c)^2 V$. The self-interaction is quadratic in $\tauh$ and can be written
\eq{
\tilde V(\tauh) & = A_{ij} \frac{\tau_{ij}}{\tau_c} + B_{ijkl} \frac{\tau_{ij} \tau_{kl}}{\tau_c^2} \\
\mathcal{\tilde V}(\tauh) & = \mathcal{A}_{ij} \frac{\tau_{ij}}{\tau_c} + \mathcal{B}_{ijkl} \frac{\tau_{ij} \tau_{kl}}{\tau_c^2}
}
We see that the left-hand-side of \eqref{RG3} will be invariant under a scaling transformation in which all lengths transform as $\ell \to \ell/b$ and all dipole moments transform as $\tau \to \tau/b^s$, provided $\mathcal{A},\mathcal{B},$ and $\mathcal{C}$ are themselves invariant.

Write 
\eq{
\Ic(\rv; b) = \frac{Z_\tau(b) a(b)^2}{\beta \tau_c(b)^2} \left\langle I \left(\frac{\rv}{b},\frac{\rvp}{b},\tauh'',\frac{\tauh}{b^s},\frac{\tauh'}{b^s} \right) \right\rangle_{\tauh''}
}
where we explicitly indicate here the dependence on $\tauh$ and $\tauh'$. If
\eq{\label{inv1}
\Ic(\rv; 1) = \Ic(\rv; b)
}
then we can make a change of variables $\rv$ to $b\rv$ in the final term in \eqref{RG3}, bringing the integration domain back to $\Omega(1)$. This term then has the original form but with modified couplings $Z_\tau(b), a(b), \tau_c(b), a(b)$ and modified variables $\tauh(b) = \tauh(1)/b^s, \tauh'(b) = \tauh'(1)/b^s$. 

We see then that the renormalized couplings will satisfy the invariance equation
\eq{
\mathcal{A}_{ij}\big[a(b),Z_\tau(b),\tau_c(b),A(b),B(b),C(b) \big] = \mathcal{A}_{ij}\big[a(1),Z_\tau(1),\tau_c(1),A(1),B(1),C(1)\big] 
}
and similarly for $\mathcal{B}$ and $\mathcal{C}$, provided the couplings renormalize as
\eq{
& \tilde V(\tauh,1) + \tilde V(\tauh',1) + \frac{a^2}{s^2} C^s_{ijkl}(1) \frac{\tau_{ij} \tau'_{kl}}{\tau_c^2} - \int_a^{ba} \frac{r dr}{a^2} \int_{\rh} \Ic(\rv; 1)- \int_a^{ba} \frac{r dr}{a^2} \int_{\hat{r}} \Ic(\vec{s}+\rv; 1) \notag \\
& \qquad = \tilde V(\tauh(b),b) + \tilde V(\tauh'(b),b) + \frac{a(b)^2}{s(b)^2} C^s_{ijkl}(b) \frac{\tau_{ij}(b) \tau'_{kl}(b)}{\tau_c(b)^2}  
}
In the limit $b \to 1^+$ we obtain
\eq{ \label{RG4}
& \frac{\p \tilde V(\tauh)}{\p \log a} + \frac{\p \tilde V(\tauh')}{\p \log a} + \frac{a^2}{s^2} \frac{\p C^s_{ijkl}}{\p \log a} \frac{\tau_{ij} \tau'_{kl}}{\tau_c^2} = - \int_{\rh} \Ic(a \rh; 1) - \int_{\hat{r}} \Ic(\vec{s}+a\rh; 1),
}
where we used the fact that $\tau_c$ transforms the same as $\tauh$ and $\tauh'$. This equation must hold for all admissible $s$, $\hat{\tau}$, and $\hat{\tau}'$; if solutions exist, then the model is renormalizable to $\OO(Z_\tau)$. 

In analogy with results for 2D melting \cite{Nelson79,Young79} the change in free energy is
\eq{
\frac{\p (\beta F)}{\p \log a} = -\half \Omega z Z_\tau^2 + \OO(Z_\tau^3) 
}
where $\Omega$ is the system volume and $z$ is the coordination number, whose precise value will not be important. 

To compute the right-hand side of \eqref{RG4}, we need to expand
\eq{
I(\rv, \rvp,\tauh'') & = e^{-\beta C^r_{ijkl} \tau_{ij} \tau{''}_{kl}/r^2} e^{-\beta C^{r'}_{ijkl} \tau'_{ij} \tau{''}_{kl}/(r')^2}-1 \notag \\
& = -\frac{\beta}{r^2} C^r_{ijkl} \tau_{ij} \tau{''}_{kl} - \frac{\beta}{(r')^2} C^{r'}_{ijkl} \tau'_{ij} \tau{''}_{kl} + \frac{\beta^2}{2r^4} \big(C^r_{ijkl} \tau_{ij} \tau{''}_{kl}\big)^2 + \frac{\beta^2}{2(r')^4} \big(C^{r'}_{ijkl} \tau'_{ij} \tau{''}_{kl} \big)^2 \notag \\
& \qquad + \frac{\beta^2}{r^2(r')^2} C^r_{ijkl} \tau_{ij} \tau{''}_{kl} C^{r'}_{pqrs} \tau'_{pq} \tau{''}_{rs} + \ldots
}
where the neglected terms will be small if the self-energy is much larger than the interaction energy. Taking the expectation over $\hat{\tau}''$, we find
\eq{
\langle I(\rv,\rvp, \tauh'') \rangle_{\tauh''} & = -\frac{\beta \tau_1}{r^2} C^r_{ijkl} \tau_{ij} \langle\tau{''}_{kl}\rangle/\tau_1 - \frac{\beta \tau_1}{(r')^2} C^{r'}_{ijkl} \tau'_{ij} \langle\tau{''}_{kl}\rangle/\tau_1 \notag \\
& + \half \beta^2 \tau_2 \left( \frac{C^r_{ijkl} \tau_{ij}}{r^2} + \frac{C^{r'}_{ijkl} \tau'_{ij}}{r'^2} \right) \frac{\langle \tau{''}_{kl} \tau{''}_{rs} \rangle}{\tau_2} \left( \frac{C^r_{pqrs} \tau_{pq}}{r^2} + \frac{C^{r'}_{pqrs} \tau'_{pq}}{r'^2} \right) + \ldots
}
where $\tau_n = \langle \tau^n \rangle$. The desired invariance property \eqref{inv1} will be satisfied if
\eq{
b^s \tau_1(b) Z_\tau(b) & = \tau_1(1) Z_\tau(1)\\
b^2 \tau_2(b) Z_\tau(b) & = \tau_2(1) Z_\tau(1) 
}
This leads to $s=1$.



There are three types of terms to consider. The first is
\eq{
I_1 & = \int_{\rh} \left\langle C^r_{ijkl} \tau_{ij} \tau{''}_{kl} \right\rangle_{\hat{\tau}''} \notag \\
& = 2\pi \tau_1 \;\tr(\tauh) (\gamma_{11}+2\lambda_{11}+\ffrac{1}{4}\lambda_{22}) + \ffrac{\pi}{8} a_\tau \gamma_{22} \;\tauh : \hat{\sigmas}/\sigmabar  
}
Note that in $I$ there is another term obtained by integrating the same function around $\rv = \vec{s} + a \rvp$; this result will be smaller by a factor of $a^2/s^2$, so we neglect it.

The second type of term is
\eq{
I_2 & = \int_{\rh} C^r_{ijkl} \tau_{ij} C^r_{pqrs} \tau_{pq} \left\langle  \tau{''}_{kl} \tau{''}_{rs} \right\rangle_{\hat{\tau}''} \\
& = \ffrac{1}{8} \tau_2 \tau_{ij} \tau_{pq} \int_{\rh} \left[ C^r_{ijkj}  C^r_{pqrq}  + 2\chi C^r_{ijkl}  C^r_{pjrl} + \OO(a_\tau) \right],
}
where in the final line we ignore corrections with a tensorial dependence on $\sigmas$, which correspond to the generation of anisotropic elasticity. After tedious algebra we find
\eq{
I_2 & = \ffrac{1}{8} \tau_2 \tau_{ik} \tau_{jl} \left[ \delta_{ik} \delta_{jl} Y_1 + Z_{ikjl} \chi Y_2 + 8\pi \chi (\delta_{ij}\delta_{kl} + \delta_{il} \delta_{jk}) Y_3 + \OO(a_\tau) \right],
}
where the $Y_i$ are as in the main text, and
\eq{
Z_{ikpr} = \int_{\rh} P^r_{ik} P^r_{pq} = \frac{\pi}{4} \left[ - \delta_{ik} \delta_{pq} + \delta_{ip} \delta_{kr} + \delta_{ir} \delta_{kp} \right]
}

Finally, we have terms of the form
\eq{
I_3 & = \int_{\rh} C^r_{ijkl} \tau_{ij} C^{r'}_{pqrs} \tau{'}_{pq} \left\langle  \tau{''}_{kl} \tau{''}_{rs} \right\rangle_{\hat{\tau}''} \\
& = \ffrac{1}{8} \tau_2 \tau_{ij} \tau{'}_{pq} \left[ \delta_{rs} \delta_{kl} + 2 \chi \delta_{jq} \delta_{ls} + \OO(a_\tau) \right] A_{ijkl} \left[ C^s_{pqrs} + \ldots \right]
}
with
\eq{
A_{ijkl} = \int_{\rh} C^r_{ijkl} = 2\pi \delta_{ik} \delta_{jl} \left[ \gamma_{11}+2\lambda_{11} + \ffrac{1}{4} \lambda_{22} \right] + \gamma_{22} Z_{ikjl}
}
After more algebra we find
\eq{
I_3 =  \ffrac{1}{8} \tau_2 \tau_{ik} \tau{'}_{jl} \left[ \half \delta_{ij} \delta_{kl} Y_4 + \delta_{ik} P^s_{jl}  Y_5 + \pi \gamma_{22} \chi C^s_{ijkl} \right]
}
and there is a related term
\eq{
I'_3 =  \ffrac{1}{8} \tau_2 \tau_{ik} \tau{'}_{jl} \left[ \half \delta_{ij} \delta_{kl} Y_4 + P^s_{ik} \delta_{jl}  Y_5 + \pi \gamma_{22} \chi C^s_{ijkl} \right]
}
Assembling terms we find
\eq{
\int_{\rh} \left\langle I(a\rv, \rvp,\tauh'') \right\rangle_{\tauh''} = -\frac{\beta}{a^2} I_1 \left[ 1 + \OO(a^2/s^2) \right] + \frac{\beta^2}{2a^4} I_2 \left[ 1 + \OO(a^4/s^4) \right] + \frac{\beta^2}{2s^2a^2} I_3 
}
and there will be another set of terms related by $\tauh \leftrightarrow \tauh'$. Comparing with $V(\tauh)$ and $C^s_{ijkl}$ this term has the correct form in order for \eqref{RG4} to be satisfied, hence the model is renormalizable under the chosen scaling ansatz. Matching up the couplings, we find the RG equations reported in the main text, where the $Y_i$ are
\eq{
Y_1 & = 8\pi(\lambda_{11}+\ffrac{1}{4}\lambda_{22}+\gamma_{11})^2+8\pi \chi(\gamma_{11}^2+\ffrac{1}{4}\gamma_{12}^2)+4\pi\chi(\lambda_{11}\lambda_{22}+\lambda_{12}^2)+16\pi\chi(\gamma_{11}\lambda_{11}+\half\gamma_{12}\lambda_{12}+\ffrac{1}{4}\gamma_{11}\lambda_{22})\\
Y_2 & = 4 (1+\chi^{-1}) (2\lambda_{12}+\gamma_{12})^2 +\gamma_{22}^2 - 16 \lambda_{12}^2 + 8 \gamma_{22} \lambda_{11} + 2 \gamma_{22}\lambda_{22} \\
Y_3 & = \lambda_{11}^2 + \half \lambda_{12}^2 + \ffrac{1}{16} \lambda_{22}^2\\
Y_4 & = 4\pi\left[4(1+\chi)(\gamma_{11}+2\lambda_{11}+\ffrac{1}{4}\lambda_{22})-\half \chi \gamma_{22}\right] (\gamma_{11}+\lambda_{11}+\ffrac{1}{4}\lambda_{22}) \\
Y_5 & = 2\pi\left[4(1+\chi)(\gamma_{11}+2\lambda_{11}+\ffrac{1}{4}\lambda_{22})-\half\chi \gamma_{22}\right] (\gamma_{12}+2\lambda_{12})
}

One check on the above computations is to see that a quadrupolar theory remains quadrupolar. Indeed, for a theory of quadrupoles $\gamma_{11}$ and $\gamma_{12}$ play no role, since they appear multiplied by $\tau_{ii}=0$. This implies that for quadrupoles, these couplings cannot appear in the $\beta$-functions of $\gamma_{22}$, $\lambda_{11}$, $\lambda_{12}$, and $\lambda_{22}$. Once the quadrupole constraint $\chi=-1$ is applied, this is indeed the case. 

{\bf Appendix 3: Derivation of dual field theory.} 

To sum over the defects, we need to first separate them using a Hubbard-Stratonovich transformation. Convergence of the associated Gaussian integral requires that we have either a positive-definite or negative-definite operator for all $\qv$. Since the original operator $\hat{F}$ is not positive-definite at all $q$, we split the interaction into an augmented operator $\tilde {\hat{F}} = \hat{F}+\delta F$ and the remainder $-\delta F$. The deformation $\delta F(\qv) = \eta k_2$ is a self-energy. We are thus shifting part of the self-energy term from directly acting on the defects to act instead on the new field introduced through the transformation.
 To be positive-definite, we need $-2\pi B \log (q/\Lambda_a) + \eta k_2 > \pi|\gamma|$ for all $q$. For $B>0$ as we will assume, the most dangerous wavenumber is $q=\Lambda$, where this reduces to $\eta > \pi|\gamma|/k_2$. 




The defects are separated using two Hubbard-Stratonovich transformations, in bra-ket notation
\eq{
e^{-\frac{1}{2} \beta \langle w-w_0 | \tilde F | w-w_0 \rangle} & \propto \int \D\zeta \; e^{-\frac{1}{2} \beta \langle \zeta | \tilde F^{-1} | \zeta \rangle} e^{i \beta \langle \zeta | w-w_0 \rangle}, \\
e^{+\frac{1}{2} \beta \langle w| \delta F | w\rangle} & \propto \int \D\epsilon' \; e^{-\frac{1}{2} \beta \langle \epsilon' | \delta F^{-1} | \epsilon' \rangle} e^{\beta \langle \epsilon' | w \rangle},
}
leading to 
\eq{
Z & = \sum_{n \geq 0} \frac{1}{n!} \int_{r_1,\ldots,r_{n}} \int_{\tau_1,\ldots,\tau_{n}} e^{-\beta H_{2n}}  \\
& = e^{\frac{1}{2} \beta \langle w_0 | \tilde F | w_0 \rangle} \sum_{n \geq 0} \frac{1}{n!} \int_{r_1,\ldots,r_{n}} \int_{\tau_1,\ldots,\tau_{n}} e^{-\frac{1}{2} \beta \langle w-w_0 | \tilde F | w-w_0 \rangle} e^{-\beta \langle w | \tilde F | w_0 \rangle} e^{+\frac{1}{2} \beta \langle w | \delta F | w \rangle}e^{-\beta \sum_a V^0(\tauh^a)} \\
& \propto |\tilde F|^{-1/2} |\delta F|^{-1/2} Z'_c \int' \D\zeta \; \int' \D\epsilon' \; \sum_{n \geq 0} \frac{1}{n!} \int_{r_1,\ldots,r_{n}} \int_{\tau_1,\ldots,\tau_{n}} e^{\beta \langle w| i \zeta +\epsilon' - \tilde F w_0\rangle} e^{-\beta \sum_a V^0(\tauh^a)}
}
where $\int' \D\zeta = \int \D\zeta \; e^{-\frac{1}{2} \beta \langle \zeta | \tilde F^{-1} | \zeta \rangle} e^{-i \beta \langle \zeta | w_0 \rangle}$ and $\int' \D\epsilon' =  \int \D\epsilon' \; e^{-\frac{1}{2} \beta \langle \epsilon' | \delta F^{-1} | \epsilon' \rangle}$. Then
\eq{
Z & \propto |\tilde F|^{-1/2} |\delta F|^{-1/2} Z'_c \int' \D\zeta \; \int' \D\epsilon' \; \sum_{n \geq 0} \frac{1}{n!} \left( \int_{r} \ffrac{1}{4\tau_c^2} \int d^2 w \; e^{\beta \vec{w} \cdot (i\zetav(\rv)-\tilde F w_0)} e^{-\beta j \vec{\sigma} \cdot \vec{w} - \frac{1}{2a^2} \beta k_2 w^2} \right)^n \\
& \propto |\tilde F|^{-1/2} |\delta F|^{-1/2} Z'_c \int' \D\zeta \; \int' \D\epsilon' \; \sum_{n \geq 0} \frac{1}{n!} \left( \ffrac{\pi a^2}{2\beta k_2 \tau_c^2} \int_{r} e^{\beta a^2 \xi(\rv)^2/(2 k_2)} \right)^n\\
& \propto |\tilde F|^{-1/2} |\delta F|^{-1/2} Z'_c \int' \D\zeta \; \int' \D\epsilon' \; \exp \left( Z_0 \int_{r} e^{\beta a^2 \xi(\rv)^2/(2 k_2)} \right)
}
where $Z'_c = e^{\frac{1}{2} \beta \langle w_0 | \tilde F | w_0 \rangle}$, $\vec{\sigma} = \sigmabar (\cos 2\theta_\sigma, \sin 2\theta_\sigma)$, $\vec{\xi}(\rv) = i\zetav(\rv) + \epsilon'(\rv) - B \log(R/a) \vec{w}_0 - \eta k_2 \wv_0 - j \vec{\sigma}$, and $Z_0 = (\pi a^2)/(2\tau_c^2 \beta k_2)$. 

It is convenient to define $\epsv_0 = B\log(R/a) \wv_0 + \eta k_2 \wv_0 + j \vec{\sigma}$, make a shift $\epsv{\;}' = \epsv + \epsv_0$, and define
\eq{
Z_c = Z_c' e^{\frac{1}{2} \beta \langle \epsv_0 | \delta F^{-1} | \epsv_0 \rangle} 
}
Assembling terms we then arrive at the nonlocal action shown in the main text.

As mentioned in the main text, the $\epsilon$ field can be eliminated in one regime. Indeed, since $\langle w | \delta F | w \rangle = \eta k_2 \sum_a \tau_a^2/a^2$, this term is just a self-energy. Instead of introducing $\epsilon$, it can be incorporated into the defect self-energy, giving a total term $\exp(-\frac{1}{2} \beta (1-\eta) w^2 k_2/a^2)$ in the $w$ integral. Convergence of this integral requires that $1-\eta>0$. Since we must have $\eta > \pi\gamma/k_2$, the former condition can be satisfied when $k_2 > \pi\gamma$, which corresponds approximately to the stable regime when $X>0$, discussed with respect to the RG. Therefore in this regime, the $\epsilon$ field is unnecessary.

\end{widetext}

\end{document}